\documentstyle[12pt]{article}
\def\zid{1\kern-0.36em\llap~1}

\newcommand{\beq}{\begin{equation}}
\newcommand{\ber}{\begin{eqnarray}}
\newcommand{\eeq}{\end{equation}}
\newcommand{\eer}{\end{eqnarray}}

\begin{document}

\begin{titlepage}
%\vbox {\vspace{0.1mm}} %Leaves space at top of 1st page.
\rightline{[SUNY BING 9/9/05 v. 3]; hep-ph/0510348}
\begin{center}
{\bf \hspace{1.85 cm} \center USE OF W-BOSON
LONGITUDINAL-TRANSVERSE INTERFERENCE \newline IN TOP QUARK SPIN-CORRELATION FUNCTIONS: II}\\
\vspace{2mm} Charles A.
Nelson\footnote{Electronic address: cnelson @ binghamton.edu  },
 Jeffrey J. Berger, and Joshua R. Wickman  \\
{\it Department of Physics, State University of New York at
Binghamton\\ Binghamton, N.Y. 13902}\\[2mm]
\end{center}

%\vspace{2mm}

\begin{abstract}

This continuation of the derivation of general beam-referenced
stage-two spin-correlation functions is for the analysis of
top-antitop pair-production at the Tevatron and at the Large
Hadron Collider. Both the gluon-production and the
quark-production contributions are included for the
charged-lepton-plus-jets reaction $p p $ or $p \bar{p} \rightarrow
t \bar{t} \rightarrow (W^+ b) (W^- \bar{b} ) \rightarrow  (l^{+}
\nu b ) (W^- \bar{b} )$. There is a simple 4-angle beam-referenced
spin-correlation function for determination of the relative sign
of or for measurement of a possible non-trivial phase between the
two dominant $\lambda_b = -1/2 $ helicity amplitudes for the
$t\rightarrow W^{+}b$ decay mode. There is an analogous function
and tests for $\bar{t} \rightarrow W^{-} \bar{b}$ decay. This
signature requires use of the $ ( t \bar{t} )_{c.m.} $ energy of
the hadronically decaying W-boson, or the kinematically equivalent
cosine of the polar-angle of $W^{\mp}$ emission in the anti-top
(top) decay frame. Spinors and their outer-products are
constructed so that the helicity-amplitude phase convention of
Jacob \& Wick can be used throughout for the fixing of the signs
associated with this large W-boson longitudinal-transverse
interference effect.
\end{abstract}

\end{titlepage}

\section{ Introduction: W-Boson Longitudinal-Transverse \newline Interference }

This continuation of a previous paper [1] on the derivation of
general beam-referenced stage-two spin-correlation functions is
for the analysis of top-antitop pair-production [2,3] at the
Tevatron and at the Large Hadron Collider [4].  Each second at the
Large Hadron Collider there will be a top-antitop pair produced
when the planned $ L \sim 10^{33} cm^{-2} sec^{-1} $ is reached.
This should provide an almost ideal ``laboratory" for both
investigating top-quark physics itself, and for simultaneously
improved empirical mastery of reaction backgrounds and detector
systematics/performance.

As in the previous paper, which we denote as ``I", the helicity
formalism [5] is used for a simple and transparent treatment of
all relative phase effects. We use this formalism for
investigating the large effects of W-boson longitudinal-transverse
interference in top-antitop pair-production for the charged-lepton
plus jets channel, the di-lepton plus jets channel, and the
all-jets channel.  In ``I", only the quark-production contribution
was included; it is the dominant contribution at the Tevatron. On
the other hand at the Large Hadron Collider, the gluon-production
contribution dominates. To leading order in $\alpha_s$, both
contributions are included in the analysis in the present paper.
The modular property of the helicity-formalism with respect to
incorporation of the various density matrices and symmetries again
remains manifest throughout this analysis.  This modularity should
be easy to exploit in understanding and checking large
interference effects in applications of these results.

Using the spinor outer-product formulas $u(p,\lambda _{1}
)\overline{u}(p,{\lambda _{1}}^{\prime} ) , \ldots $ obtained in
Appendix A, the associated gluon production
density-matrix-elements are derived in Appendix B in the
helicity-amplitude phase convention of Jacob \& Wick (JW).  The
analogous quark production density-matrix-elements were obtained
in ``I". They also follow from these spinor outer-products.
Besides their dependence on $\cos\Theta_B$, these gluon and quark
density-matrix-elements $ \rho _{\lambda _1\lambda _2,\lambda
_1^{^{\prime }}\lambda _2^{^{\prime }}}(\Theta _B,\Phi _B)$
exhibit several non-trivial overall minus signs and an explicit
dependence on the azimuthal-angle $\Phi_B$. These important
spherical angles $\Theta_B$ and $\Phi_B$ are defined in Figs. 1
and 2. Otherwise, to avoid repetition, we assume that the reader
has ``I" available with its figures, discussions, and the
sequential-decay density matrices $R_{\lambda _1\lambda
_1^{^{\prime }}} (t\rightarrow W^{+}b\rightarrow (l^{+}\nu )b )$
and $\overline{ { R } }_{\lambda _2\lambda _2^{^{\prime }}}(\bar
t\rightarrow W^{-}\bar b\rightarrow ( l^- \bar{\nu} ) \bar{b} )$
for the $CP$-conjugate process. For the sequential-decay with the
first-stage $ t \rightarrow W^{+} b $ followed by the second-stage
$ W^+ \rightarrow l^+ \nu $, the spherical angles $ \theta_a$,
$\phi_a $ specify the $ l^+ $ momentum in the ${W_1}^+$ rest frame
when there is first a boost from the $ (t \bar{t} )_{c.m.}$ frame
to the $t_1$ rest frame, and then a second boost from the $t_1$
rest frame to the ${W_1}^+$ rest frame ( see ``I").   The $0^o$
direction for the azimuthal-angle $\phi_a$ is defined by the
projection of the ${W_2}^-$ momentum direction. Analogously, the
spherical angles $ \theta_b$, $\phi_b $ specify the $ l^- $
momentum in the ${W_2}^-$ rest frame.

As in ``I", the emphasis is on tests for the
charged-lepton-plus-jets reaction $p p$ or $p \bar{p} \rightarrow
t \bar{t} \rightarrow (W^+ b) (W^- \bar{b} ) \rightarrow  (l^{+}
\nu b ) (W^- \bar{b} )$.  However, in contrast with the analysis
in ``I", because of the differences in dependence on
$\cos\Theta_B$ among the 5 sets of gluon-production
density-matrix-elements, for clarity $\cos\Theta_B$ is not
integrated out in the present paper. Consequently, versus the
3-angle stage-two spin-correlation function ${ \mathcal{F}{|}}
_{0} +
 {\mathcal{F}{|}}_{sig} $ in ``I" which only included the
quark-production contribution, there is instead a simple 4-angle
beam-referenced stage-two spin-correlation function ${
\mathcal{G}}^{(g,q)} {|} _{0} + {\mathcal{G}}^{(g,q)} {|}_{sig} $
with both gluon (2,3) and quark (9,10) production contributions.

This BR-S2SC function can be used for determination of the
relative sign of or for measurement of a possible non-trivial
phase between the two dominant $\lambda_b = -1/2 $ helicity
amplitudes for the $t\rightarrow W^{+}b$ decay mode. Both in the
SM and in the case of an additional large $t_R \rightarrow b_L $
chiral weak-transition moment [6], the $\lambda _{b}=-1/2$ and $
\lambda \overline{_{b}}$ $=1/2$ amplitudes are more than $\sim 30$
times larger than the $\lambda _{b}=1/2$ and $\lambda
\overline{_{b}}$ $=-1/2$ amplitudes. For the $CP$-conjugate case,
there are analogous tests for $\bar{t} \rightarrow W^{-} \bar{b}$
decay based on the analogous function, (21,22) and (24,25).

As in ``I", this important signature from W-boson
longitudinal-transverse interference requires use of the $ ( t
\bar{t} )_{c.m.} $ energy of the hadronically decaying W-boson, or
the kinematically equivalent cosine of the polar-angle of
$W^{\mp}$ emission in the anti-top (top) decay frame, $\cos{\theta
_{2,1}^{t}}$. In application, for instance to $p p \rightarrow t
\bar{t} X$, parton-level top-quark spin-correlation functions
[7,8] need to be smeared with the appropriate parton-distribution
functions with integrations over $\cos\Theta_B$ and the $ ( t
\bar{t} )_{c.m.} $ energy, $\sqrt{s}$.  The color factors have
been included in these BR-S2SC functions.

For the gluon-production-decay sequence
\begin{equation}
g_1 g_2 \rightarrow
t_1 \overline{t}_2 \rightarrow (W^{+}b)(W^{-}%
\overline{b}) \rightarrow \cdots ,
\end{equation}
we assume that the $\lambda _{b}=-1/2$ and $\lambda \overline{_{b}}$ $%
=1/2$ amplitudes dominate in $t_1$ and $\bar{t_2}$ decay. The
simple four-angle distribution ${ \mathcal{G}}^{g}{|} _{0} +
 {\mathcal{G}}^{g}{|}_{sig} $ for $%
{t}_{1}\rightarrow W_{1}^{+}{b}\rightarrow ( l^{+}\nu ) {b}$
involves the angles \newline $\{ \Theta_B$; $\theta _{2}^{t}$, $
\theta _{a}$, $\phi _{a} \}$, where $ \Theta_B$ is shown in Fig.1.
The other angles are displayed in Figs. 1 and 3 of ``I".  The
``gluon contribution" is
\begin{eqnarray}
{ \mathcal{G}}^{g}{|} _{0} &=& \frac{16\pi ^{3}}{3} \,\,
c(s,\Theta
_{B}) \,\, \widetilde{g}_{1}^{g}(s,\Theta _{B}) \nonumber \\
&&\left\{ \frac{1}{2}\Gamma (0,0)\sin ^{2}\theta _{a}+\Gamma (-1,-1)\sin ^{4}%
\frac{\theta _{a}}{2}\right\} [\overline{\Gamma }(0,0)+\overline{\Gamma }%
(1,1)]
\end{eqnarray}
\begin{eqnarray}
{\mathcal{G}}^{g}{|}_{sig} &=&{-}\frac{4\sqrt{2}\pi ^{4}}{3}
\,\,c(s,\Theta _{B}) \,\, \widetilde{g}_{2}^{g}(s,\Theta _{B})
\,\, \cos \theta _{2}^{t}\sin \theta _{a}\sin ^{2}\frac{\theta
_{a}}{2}
\nonumber \\
&&\left\{
\Gamma _{R}(0,-1)\cos \phi _{a}-\Gamma _{I}(0,-1)\sin \phi _{a}\right\} [%
\overline{\Gamma }(0,0)+\overline{\Gamma }(1,1)]{\mathcal{R}}
\end{eqnarray}
with two gluon-beam-referencing factors
\begin{eqnarray}
\widetilde{g}_{1}^{g}(s,\Theta _{B}) &=&\sin ^{2}\Theta
_{B}(1+\cos
^{2}\Theta _{B})+\frac{8m^{2}}{s}(\cos ^{2}\Theta _{B}+\sin ^{4}\Theta _{B})-%
\frac{16m^{4}}{s^{2}}(1+\sin ^{4}\Theta _{B}) \\
\widetilde{g}_{2}^{g}(s,\Theta _{B}) &=&\sin ^{2}\Theta
_{B}(1+\cos
^{2}\Theta _{B})-\frac{8m^{2}}{s}( 1 +\sin ^{2}\Theta _{B})+%
\frac{16m^{4}}{s^{2}}(1+\sin ^{4}\Theta _{B})
\end{eqnarray}
The tildes denote the fact that these two factors first appear in
the slightly more general five-angle BR-S2SC functions in Sec. 2
in which the $(t \bar{t})_{c.m.}$ energy of the leptonically
decaying $W^{+}$ has not been integrated out, i.e. in which
$\cos{{\theta_1}^t}$ dependence has not been integrated out. The
overall pole-factor
\begin{equation}
c(s,\Theta _{B})=\frac{ s^{2} g^4 }{96 (m^{2}-t)^{2}(m^{2}-u)^{2}} \,\, [7+%
\frac{36p^{2}}{s}\cos ^{2}\Theta _{B}]
\end{equation}
depends on the $(t \bar{t})_{c.m.}$ center-of-momentum energy
$\sqrt{s}$ and $\cos{\Theta_B}$, and includes the gluon color
factor. The ``signal" contribution is suppressed by
\begin{equation}
 {\mathcal{R}} \equiv \frac{[{\overline{\Gamma
}}(0,0)-{\overline{\Gamma }}(1,1)]} {[{\overline{\Gamma
}}(0,0)+{\overline{\Gamma }}(1,1)]}
\end{equation}
which is associated with the stage-one part of the sequential
decay $\bar{t} \rightarrow W^-  \bar{b}$.  There is similarly a
suppression factor $ \overline{{\mathcal{R}}} $,  see (23),  for
the CP-conjugate
channel $\overline{t}%
_{2}\rightarrow W_{2}^{-}\overline{b}\rightarrow
(l^{-}\bar{\nu})\overline{b} $. Numerically, ${\mathcal{R}} =
\overline{{\mathcal{R}}} = (1-\frac{2m_{W}^{2}}{{m_t}^{2}}) /
(1+\frac{2m_{W}^{2}}{{m_t}^{2}}) \sim 0.41$ in both the standard
model and in the case of an additional large $t_R \rightarrow b_L$
chiral weak-transition moment.

For $g_{1}(q)g_{2}(r)\longrightarrow t_{1}(p)\overline{t_{2}}(l)$,
in terms of the particles' external momenta, useful kinematic
formulas are: $\
s=(q+r)^{2}=(p+l)^{2}=4E^{2},t=(p-q)^{2}=(r-l)^{2}=m^{2}-2E^{2}(1-\beta
\cos \Theta _{B}),$ and
$u=(p-r)^{2}=(q-l)^{2}=m^{2}-2E^{2}(1+\beta \cos \Theta _{B})$
where $\beta =p/E,\gamma =E/m$ with $m, E, p$ the top-quark mass,
energy, and magnitude of 3-momentum in the $(t\bar{t})_{c.m.}$
frame.

For the quark-production-decay sequence
\begin{equation}
q_1 \overline{q_2} \rightarrow
t_1 \overline{t}_2 \rightarrow (W^{+}b)(W^{-}%
\overline{b}) \rightarrow \cdots ,
\end{equation}
we also assume that the $\lambda _{b}=-1/2$ and $\lambda \overline{_{b}}$ $%
=1/2$ amplitudes dominate in $t_1$ and $\bar{t_2}$ decay.
Including the color factor, the four-angle distribution ${
\mathcal{G}}^{q}{|} _{0} +
 {\mathcal{G}}^{q}{|}_{sig} $ for $%
{t}_{1}\rightarrow W_{1}^{+}{b}\rightarrow ( l^{+}\nu ) {b}$ or
``quark contribution" is
\begin{eqnarray}
{\mathcal{G}}^{q}{|}_{0}{} &=&{}\frac{2\pi ^{3}g^{4}}{27 s^{2}}
\,\, \widetilde{g}_{1}^{q}(s,\Theta _{B})
\nonumber \\
&&\left\{ \frac{1}{2}\Gamma (0,0)\sin ^{2}\theta _{a}+\Gamma (-1,-1)\sin ^{4}%
\frac{\theta _{a}}{2}\right\} [\overline{\Gamma }(0,0)+\overline{\Gamma }%
(1,1)]
\end{eqnarray}
\begin{eqnarray}
{\mathcal{G}}^{q}{|}_{sig} &=&{-}\frac{\sqrt{2}\pi ^{4}g^{4}}{54 s^{2}} \,\,%
\widetilde{g}_{2}^{q}(s,\Theta _{B}) \,\,
\cos \theta _{2}^{t}\sin
\theta _{a}\sin ^{2}\frac{\theta _{a}}{2}
\nonumber \\
&&\left\{
\Gamma _{R}(0,-1)\cos \phi _{a}-\Gamma _{I}(0,-1)\sin \phi _{a}\right\} [%
\overline{\Gamma }(0,0)+\overline{\Gamma }(1,1)]{\mathcal{R}}
\end{eqnarray}
with two quark-beam-referencing factors
\begin{eqnarray}
\widetilde{g}_{1}^{q}(s,\Theta _{B}) &=&1+\cos ^{2}\Theta _{B}+\frac{4m^{2}}{%
s}\sin ^{2}\Theta _{B} \\
\widetilde{g}_{2}^{q}(s,\Theta _{B}) &=&1+\cos ^{2}\Theta _{B}-\frac{4m^{2}}{%
s}\sin ^{2}\Theta _{B}
\end{eqnarray}

In terms of the $t\rightarrow W^{+}b$ helicity amplitudes $A\left(
\lambda _{W^{+} } ,\lambda _b \right)$ in the $t_1$ rest frame,
the decay density matrix element
\begin{eqnarray}
\langle \theta _1^t ,\phi _1 ,\lambda _{W^{+} } ,\lambda _b |\frac
12,\lambda _1\rangle =D_{\lambda _1,\mu }^{(1/2)*}(\phi _1 ,\theta
_1^t ,0)A\left( \lambda _{W^{+} } ,\lambda _b \right), \\
A(\lambda _{W^{+}},\lambda _{b})\equiv \left| A(\lambda
_{W^{+}},\lambda _{b})\right| \exp (\imath \, \, \varphi _{\lambda
_{W^{+}},\lambda _{b} }) \nonumber
\end{eqnarray}
where $\mu =\lambda _{W^{+} } -\lambda _b $ in terms of the
${W_1}^+$ and $b$-quark helicities.  The dominant
polarized-partial-widths are
\begin{eqnarray}
\Gamma (0,0) & \equiv &\left| A(0,-1/2)\right| ^{2}, \, \, \Gamma
(-1,-1)\equiv \left| A(-1,-1/2)\right| ^{2}
\end{eqnarray}
and W-boson-LT-interference-widths are
\begin{eqnarray}
\Gamma _{\mathit{R}}(0,-1) &=&\Gamma _{\mathit{R}}(-1,0)\equiv {{
Re}[A(0,-1/2)A(-1,-1/2)^{\ast }}]  \nonumber \\
& \equiv & |A(0,-1/2)||A(-1,-1/2)|\cos \beta _{L}  \\
\Gamma _{\mathit{I}}(0,-1) &=&-\Gamma _{\mathit{I}}(-1,0) \equiv {Im}%
[A(0,-1/2)A(-1,-1/2)^{\ast }]  \nonumber \\
& \equiv &-|A(0,-1/2)||A(-1,-1/2)|\sin \beta _{L}
\end{eqnarray}
where $\beta _{L}  \equiv  \varphi _{-1,-\frac{1}{2}}-\varphi
_{0,-\frac{1}{2}}$.

For the $CP$-conjugate process, $\bar t_2 \rightarrow {W_2}^{-}
\bar b$, in the $\bar t_2$ rest frame, the helicity amplitudes
$B\left( \lambda _{W^{-} },\lambda _{\bar b}\right)$ are defined
by
\begin{eqnarray}
\langle \theta _2^t ,\phi _2 ,\lambda _{W^{-} },\lambda _{\bar
b}|\frac 12,\lambda _2\rangle =D_{\lambda _2,\bar \mu
}^{(1/2)*}(\phi _2 ,\theta _2^t ,0)B\left( \lambda _{W^{-}
},\lambda _{\bar b}\right), \\
B(\lambda _{W^{-}},\lambda _{\overline{b}})\equiv \left| B(\lambda
_{W^{-}},\lambda \overline{_{b}})\right| \exp (\imath \, \,
\overline{\varphi }_{\lambda _{W^{-}},\lambda \overline{_{b}}})
\nonumber
\end{eqnarray}
with $\bar \mu =\lambda _{W^{-}}-\lambda _{\bar b}$, and
\begin{eqnarray}
\overline{\Gamma }(0,0) & \equiv &\left| B(0,1/2)\right| ^{2}, \, \, \overline{%
\Gamma }(1,1)\equiv \left| B(1,1/2)\right| ^{2}  \\
\overline{\Gamma }_{R}(0,1) &=&\overline{\Gamma }_{\mathit{R}%
}(1,0)\equiv {{Re}[B(0,1/2)B(1,1/2)^{\ast }}]  \nonumber \\
& \equiv &|B(0,1/2)||B(1,1/2)|\cos \overline{\beta }_{R}  \\
\overline{\Gamma }_{I}(0,1) &=& - \overline{\Gamma }_{I}(1,0)
 \equiv {{Im} [B(0,1/2)B(1,1/2)^{\ast }}]  \nonumber \\
& \equiv &-|B(0,1/2)||B(1,1/2)|\sin \overline{\beta }_{R}
\end{eqnarray}
where $\overline{\beta }_{R}  \equiv \overline{\varphi }_{1,\frac{1}{2}}-\overline{%
\varphi }_{0,\frac{1}{2}}$.

For the $CP$-conjugate channel $%
\overline{t}_{2}\rightarrow W_{2}^{-}\overline{b}\rightarrow (
l^{-}\bar{\nu} ) \overline{b}$, the analogous four-angle BR-S2SC
function $ \overline{{ \mathcal{G}}^{g} {|}} _{0} +
\overline{{\mathcal{G}}^{g} {|}}_{sig} $  is a distribution versus
$\{ \Theta_B$; $\theta _{1}^{t}$, $ \theta _{b}$, $\phi _{b} \}$,
where the latter three angles are displayed in in Figs. 2 and 3 of
``I". The gluon contribution is
\begin{eqnarray}
\overline{{ \mathcal{G}}^{g} {|}} _{0} &=&  \frac{16\pi ^{3}}{3}
\,\,c(s,\Theta _{B}) \,\, \widetilde{g}_{1}^{g}(s,\Theta _{B})
\nonumber \\
&&\left\{ \frac{1}{2}\overline{\Gamma }(0,0)\sin ^{2}\theta _{b}+\overline{%
\Gamma }(1,1)\sin ^{4}\frac{\theta _{b}}{2}\right\} [\Gamma
(0,0)+\Gamma (-1,-1)]
\end{eqnarray}
\begin{eqnarray}
\overline{{\mathcal{G}}^{g} {|}}_{sig} &=&{-}\frac{4\sqrt{2}\pi ^{4}}{3} \,\,%
c(s,\Theta _{B})  \,\, \widetilde{g}_{2}^{g}(s,\Theta _{B}) \,\,
\cos \theta _{1}^{t}\sin \theta _{b}\sin ^{2}\frac{\theta _{b}}{2}
\nonumber \\
&&\left\{ \overline{\Gamma }_{R}(0,1)\cos \phi
_{b}+\overline{\Gamma }_{I}(0,1)\sin \phi _{b}\right\} [\Gamma
(0,0)+\Gamma (-1,-1)]\overline{{\mathcal{R}}}
\end{eqnarray}
where
\begin{equation}
\overline{{\mathcal{R}}} \equiv  \frac{[{\Gamma }(0,0)-{\Gamma
}(-1,-1)]} {[{\Gamma }(0,0)+{\Gamma }(-1,-1)]}
\end{equation}
The quark contribution is
\begin{eqnarray}
\overline{{ \mathcal{G}}^{q} {|}} _{0} &=&  \frac{2\pi ^{3}
g^4}{27 s^2}  \,\, \widetilde{g}_{1}^{q}(s,\Theta _{B})
\nonumber \\
&&\left\{ \frac{1}{2}\overline{\Gamma }(0,0)\sin ^{2}\theta _{b}+\overline{%
\Gamma }(1,1)\sin ^{4}\frac{\theta _{b}}{2}\right\} [\Gamma
(0,0)+\Gamma (-1,-1)]
\end{eqnarray}
\begin{eqnarray}
\overline{{\mathcal{G}}^{q} {|}}_{sig} &=&{-}\frac{\sqrt{2}\pi ^{4}g^4}{54 s^2}%
 \,\, \widetilde{g}_{2}^{q}(s,\Theta _{B}) \,\,
 \cos \theta _{1}^{t}\sin \theta _{b}\sin ^{2}\frac{\theta
_{b}}{2}
\nonumber \\
&&\left\{ \overline{\Gamma }_{R}(0,1)\cos \phi
_{b}+\overline{\Gamma }_{I}(0,1)\sin \phi _{b}\right\} [\Gamma
(0,0)+\Gamma (-1,-1)]\overline{{\mathcal{R}}}
\end{eqnarray}

\subsection{Structure of four-angle BR-S2SC functions}

In general, the $t_1 \overline{t}_2 \rightarrow (W^{+}b)(W^{-}%
\overline{b}) \rightarrow (l^{+} \nu b)(W^{-} \overline{b}) $
decay-structure of the above four-angle BR-S2SC functions is
exactly analogous to that of the quark-production three-angle
non-beam-referenced S2SC functions in ``I".  The significant
difference is the additional dependence on $\cos{\Theta_B}$ due to
the beam-referencing.  Therefore, for gluon-production there are
the two gluon-beam-referencing factors
$\widetilde{g}_{1}^{g}(s,\Theta _{B}) $,
$\widetilde{g}_{2}^{g}(s,\Theta _{B}) $ of (4,5), and for
quark-production the two quark-beam-referencing factors
$\widetilde{g}_{1}^{q}(s,\Theta _{B}) $,
$\widetilde{g}_{2}^{q}(s,\Theta _{B}) $ of (11,12).  There is a
common final-state interference structure in these BR-S2SC
functions for the charged-lepton plus jets reaction $p p $ or $p
\bar{p} \rightarrow t \bar{t} \rightarrow \ldots $.  The
final-state relative phase effects do not depend on whether the
final $t_1 \overline{t}_2$ system has been produced by gluon or by
quark production.

In these four-angle expressions, the signal contributions are
again suppressed by the factor ${\mathcal{R}} = ({\mathtt{prob}}
\, W_L) - ({\mathtt{prob}} \, W_T) $ as a consequence of the
dynamical assumption that the
$\lambda _{b}=-1/2$ and $\lambda \overline{_{b}}$ $%
=1/2$ amplitudes dominate. From the ${\theta_2}^t$ dependence of
the integrated diagonal-elements of the sequential-decay density
matrices $\overline{\mathbf{R}}_{++}^{\overline{b}_{R}}$,
$\overline{\mathbf{R}}_{--}^{\overline{b}_{R}}$ for $\bar{t_2}
\rightarrow {W_2}^- \bar{b} \rightarrow ( l^- \bar{\nu} ) \bar{b}
$, $\overline{\mathbf{R}}_{++}^{\overline{b}_{R}}$ and
$\overline{\mathbf{R}}_{--}^{\overline{b}_{R}}$  [ see (95-96) of
``I"], it follows that ${\mathcal{R}}$'s numerator appears in
${\mathcal{G}}^{g,q}{|}_{sig} $ multiplied by $\cos \theta
_{2}^{t}$ and that ${\mathcal{R}}$'s denominator appears in
${\mathcal{G}}^{g,q}{|}_{0} $ multiplied by one. The off-diagonal
$\overline{R} _{\lambda _2\lambda _2^{^{\prime }}}$ elements which
describe $\bar{t_2}$-helicity interference do not contribute due
to the integration over the opening-angle $\phi$ between the $t_1$
and $\bar{t_2}$ decay planes.  There are the analogous structures
in the four-angle $ \overline{{ \mathcal{G}}^{g,q} {|}} _{0} +
\overline{{\mathcal{G}}^{g,q} {|}}_{sig} $ functions for the
CP-conjugate tests.

\subsection{Outline of This Paper}

For the production-decay sequence $g_1 g_2 \rightarrow
t_1 \overline{t}_2 \rightarrow (W^{+}b)(W^{-}%
\overline{b}) \rightarrow \cdots $, Section 2 of this paper
contains the derivation in the $ (t \bar{t} )_{c.m.}$ frame of the
gluon-production BR-S2SC functions in the ``home" or starting
coordinate system $ ( {x_h}, {y_h}, {z_h} ) $, which is defined by
Figs. 1 and 2 with the top-quark $t_1$ is moving along the
positive $z_h$ direction.  Sec. 2.1 lists the gluon-production
$t_{1}\overline{t_{2}}$ density-matrix-elements in this coordinate
system with the $g_1$ gluon-momentum ``beam" direction specified
by the spherical angles $\Theta_B, \Phi_B$. In Sec. 2.2, the gluon
contributions to the general BR-S2SC functions $I _{\lambda
_1\lambda _2;\lambda _1^{^{\prime }}\lambda _2^{^{\prime }}}^{g}$
are obtained, and in Sec. 2.3 applied to the lepton-plus-jets
channel of the $t \bar{t}$ system, assuming that the $\lambda_b =
- 1/2$ and $\lambda_{\bar{b}} = 1/2$ amplitudes dominate.  The
above four-angle gluon-production BR-S2SC function ${
\mathcal{G}}^{g}{|} _{0} +
 {\mathcal{G}}^{g}{|}_{sig} $ is obtained, along with
addition-angle generalizations which might be useful empirically.
Analogously, in Section 3, the quark-production
$t_{1}\overline{t_{2}}$ density-matrix-elements for Figs. 1 and 2
are listed with $q_1$ quark-momentum ``beam" direction specified
by the spherical angles $\Theta_B, \Phi_B$. The analogous BR-S2SC
functions are obtained for the case of quark-production  $q_1
\overline{q_2} \rightarrow
t_1 \overline{t}_2 \rightarrow (W^{+}b)(W^{-}%
\overline{b}) \rightarrow \cdots $.  Section 4 contains a summary
and a discussion.  The appendices respectively contain (A) the
Dirac spinors and their outer-products $u(p,\lambda _{1}
)\overline{u}(p,{\lambda _{1}}^{\prime} ), \ldots $ in the JW
phase convention including the C-P-T discrete symmetry properties
of the spinors, (B) the derivation of the gluon-production
density-matrix-elements in the JW phase convention for Figs. 1 and
2, and (C) the alternative $\Theta_t$, $\Phi_t$ production
density-matrix-elements for the alternative beam-reference system
defined by Fig. 3, in which the incident parton ``beam" specifies
the positive $z_b$ axis.

\section{Derivation of Gluon-Production Beam-Referenced \newline Stage-Two
 Spin-Correlation Functions}

The general beam-referenced angular distribution in the $(t\bar
t)_{cm}$ is
\begin{equation}
\begin{array}{c}
I(\Theta _B,\Phi _B;\theta _1^t,\phi _1; {\theta _a},{\phi
_a};\theta _2^t,\phi
_2;{%
\theta _b},{\phi _b}) = \sum_{\lambda _1\lambda _2\lambda
_1^{^{\prime }}\lambda _2^{^{\prime }}}\rho _{\lambda _1\lambda
_2,\lambda _1^{^{\prime }}\lambda _2^{^{\prime
}}}^{\mathtt{prod}}(\Theta _B,\Phi _B) \\
\times R_{\lambda _1\lambda _1^{^{\prime }}} (t\rightarrow
W^{+}b\rightarrow \ldots )
 \overline{ { R } }_{\lambda _2\lambda
_2^{^{\prime }}}(\bar t\rightarrow W^{-}\bar b\rightarrow \ldots )
\end{array}
\end{equation}
where the summations are over the $t_1$ and $\bar{t}_2$
helicities.  The composite decay-density-matrices $R_{\lambda
_1\lambda _1^{^{\prime }}}$ for $t\rightarrow W^{+}b\rightarrow
\ldots $ and ${\overline{R}_{\lambda _2\lambda _2^{^{\prime }}}}$
for $\bar t\rightarrow W^{-}\bar b\rightarrow \ldots $ are given
in Sec. 2.1 of ``I". This formula holds for any of the above $t
\bar{t}$ production channels and for either the lepton-plus-jets,
the dilepton-plus-jets, or the all-jets $t \bar{t}$ decay
channels.

In the $(t\bar{t})_{c.m.}$ frame, Figs. 1 and 2, the angles
$\Theta _B,\Phi _B$ specify the direction of the incident beam,
the $g_1$ or $q_1$ momentum. The $t_1$ momentum is chosen to lie
along the positive $z_h$ axis. The positive $x_h$ direction is an
arbitrary, fixed perpendicular direction. Because the incident
beam is assumed to be unpolarized, there is no dependence on the
associated $\phi_1$ angle after the observable azimuthal-angles
are specified (see below). With respect to the normalization of
the various BR-S2SC functions, the $\phi_1$ integration is not
explicitly performed in this paper. With (26) there is an
associated differential counting rate
\begin{equation}
\begin{array}{c}
dN=I(\Theta _B,\Phi _B;\ldots )d(\cos \Theta _B)d\Phi _B d(\cos
\theta _1^t)d\phi _1 \\
d(\cos {\theta _a})d{\phi _a}d(\cos \theta _2^t)d\phi _2 d(\cos
{\theta _b})d{\phi _b}
\end{array}
\end{equation}
where, for full phase space, the cosine of each polar-angle ranges
from -1 to 1, and each azimuthal-angle ranges over $ 2 \pi $.

We use the helicity indices to label the successive terms in the
gluon contributions to the sum in (26).  Each term is denoted $I
_{\lambda _1\lambda _2;\lambda _1^{^{\prime }}\lambda _2^{^{\prime
}}}^{g}$ with $\lambda _1\lambda _2, \ldots$ the signs of the
$t_1$,$\bar{t_2}$ helicities.  Unlike in ``I", in this paper we do
not use superscripts ``m" and ``m2" to emphasize the
mixed-helicity and helicity-flip contributions versus the
non-superscripted helicity-conserving ones, since a quick glance
at the patterns in these subscripted helicity indices provides
these distinctions. The charged lepton's azimuthal-angle $\phi_a$,
or $\phi_b$ is always referenced by the opposite $W^{\mp}$-boson
momentum, so these charged-lepton azimuthal-angles are denoted
without ``tilded accents".   For the alternative
$\widetilde{\phi_a}$ and $\widetilde{\phi_b}$ referencing by the
opposite $\bar{t_2}$, $t_1$ momentum directions, see the
discussion in 2nd and 3rd paragraphs of Section 2 of ``I".

\subsection{ Gluon-production density matrix in Jacob and Wick phase convention}

The $t_{1}\overline{t}_{2}$ helicity-conserving gluon-production
density-matrix-elements in the $(t\overline{t})_{c.m.}$ system are
\begin{equation}
\rho _{+-,+-}^{g}=\rho _{-+,-+}^{g}= c(s,\Theta _{B}) \,\,
\frac{4p^{2}}{s} \,\,  \sin ^{2}\Theta _{B}(1+\cos ^{2}\Theta
_{B})
\end{equation}
\begin{equation}
\rho _{-+,+-}^{g}=\{\rho _{+-,-+}^{g}\}^{\ast }=c(s,\Theta _{B})
\,\, \frac{4p^{2}}{s} \,\, e^{i2\Phi _{B}} \,\, \sin ^{4}\Theta
_{B}
\end{equation}
where the asterisk denotes complex-conjugation, and the common
pole-factor $c(E,\Theta_B)$ is given in (6) in the introduction.
Note the $ e^{i 2 \Phi_{B}} $ factor in (29).

The mixed helicity-properties gluon-production
density-matrix-elements are
\begin{eqnarray}
\rho _{-+,++}^{g} &=&\rho _{-+,--}^{g}=-\rho _{--,+-}^{g}=-\rho
_{++,+-}^{g}=
\\
\{\rho _{++,-+}^{g}\}^{\ast } &=&\{\rho _{--,-+}^{g}\}^{\ast
}=-\{\rho
_{+-,--}^{g}\}^{\ast }=-\{\rho _{+-,++}^{g}\}^{\ast }= \\
&&-c(s,\Theta _{B}) \,\, \frac{8p^{2}m}{s^{3/2}} \,\, e^{i\Phi
_{B}} \,\, \sin ^{3}\Theta _{B}\cos \Theta _{B}
\end{eqnarray}
with an overall minus sign and $e^{i\Phi _{B}}$ factor.

The helicity-flip gluon-production density-matrix-elements are
\begin{equation}
\rho _{++,++}^{g}=\rho _{--,--}^{g}=c(s,\Theta _{B}) \,\,
\frac{4 m^{2}}{s} \,\, (1+%
\frac{4 p^{2}}{s} \,\, [1+\sin ^{4}\Theta _{B}])
\end{equation}
\begin{equation}
\rho _{++,--}^{g}=\rho _{--,++}^{g}=c(s,\Theta _{B}) \,\,
\frac{4 m^{2}}{s} \,\,(1-%
\frac{4p^{2}}{s} \,\, [1+\sin ^{4}\Theta _{B}])
\end{equation}

\subsection{Gluon contributions to BR-S2SC function}

The helicity-conserving contribution is
\begin{eqnarray}
I_{+-,+-}^{g}=  \frac{4p^2}{s} c(s,\Theta
_{B}) \,\, {\mathbf{R}_{++}} \overline{{\mathbf{R}}}_{--}%
\sin ^{2} \Theta _{B}  ( 1 + \cos ^{2} \Theta _{B}   ) \\
I_{-+,-+}^{g}=  \frac{4p^2}{s} c(s,\Theta
_{B}) \,\,  {\mathbf{R}_{--}} \overline{{\mathbf{R}}}_{++}%
\sin ^{2} \Theta _{B}  ( 1 + \cos ^{2} \Theta _{B}  )  \\
I_{+-,-+}^{g}=  \frac{4p^2}{s} c(s,\Theta
_{B}) \,\,  e^{-\imath (2\Phi _{R}+\phi )} \,\, %
{\mathbf{r}_{+-}} \overline{{\mathbf{r}}}_{-+}  \sin
^{4}\Theta _{B}   \\
I_{-+,+-}^{g}=  \frac{4p^2}{s} c(s,\Theta
_{B}) \,\,  e^{\imath (2\Phi _{R}+\phi )} \,\, %
{\mathbf{r}_{-+}} \overline{{\mathbf{r}}}_{+-}  \sin ^{4}\Theta
_{B}
\end{eqnarray}
where ${\mathbf{R}_{++}}$, ${\mathbf{R}_{--}}$,
$\overline{{\mathbf{R}}}_{++}$, $\overline{{\mathbf{R}}}_{--}$,
are the diagonal elements, and $({\mathbf{r}_{-+}})^* =
{\mathbf{r}_{+-}} =F_{a}+\imath H_{a}$,
$({\mathbf{\overline{r}_{-+}}})^*= {\mathbf{\overline{r}_{+-}}} =
-F_{b}-\imath H_{b}$ the off-diagonal elements of the
sequential-decay density matrices $R_{\lambda _1\lambda
_1^{^{\prime }}} (t\rightarrow W^{+}b\rightarrow (l^{+}\nu )b )$
and $\overline{ { R } }_{\lambda _2\lambda _2^{^{\prime }}}(\bar
t\rightarrow W^{-}\bar b\rightarrow ( l^- \bar{\nu} ) \bar{b} )$.
As in Sec. 2.2.1 of ``I", in the $ { I^{g} }_{\lambda _1\lambda
_2,{ \lambda _1}^{\prime} {\lambda _1}^{\prime}}  $ and $ { I^{q}
}_{\lambda _1\lambda _2,{ \lambda _1}^{\prime} {\lambda
_1}^{\prime}}  $ contributions in this paper, the starting angles
$\phi_2$ and $\Phi_B$ have been replaced by the angles $ \phi =
\phi_1 + \phi_2 $ and $ \Phi_R = \Phi_B - \phi_1 $.

The first part of the mixed helicity-properties contribution is
\begin{eqnarray}
I_{++,+-}^{g}= -  \frac{8p^2 m}{s^{3/2}}  \,\, c(s,\Theta _{B})
\,\,
 {\mathbf{R}}_{++}(F_{b}+\imath H_{b}) \sin^{3} \Theta _{B} \cos
\Theta _{B} e^{ \imath ( \Phi _{R} + \phi )}
\end{eqnarray}
\begin{eqnarray}
I_{--,-+}^{g}=  \frac{8p^2 m}{s^{3/2}}  \,\, c(s,\Theta _{B}) \,\,
{\mathbf{R}}_{--}(F_{b}-\imath H_{b}) \sin^{3} \Theta _{B} \cos
\Theta _{B} e^{- \imath ( \Phi _{R} + \phi )}
\end{eqnarray}
\begin{eqnarray}
I_{++,-+}^{g}= -  \frac{8p^2 m}{s^{3/2}} \,\,  c(s,\Theta _{B})
\,\, {(F_{a}+\imath H_{a})}
 \overline{{{\mathbf{R}%
}}}{_{++}} \sin^{3} \Theta _{B} \cos \Theta _{B}
 e^{-
\imath \Phi _{R} }
\end{eqnarray}
\begin{eqnarray}
I_{--,+-}^{g}=   \frac{8p^2 m}{s^{3/2}} \,\, c(s,\Theta _{B}) \,\,
{(F_{a}-\imath H_{a})}
 \overline{{{\mathbf{R}%
}}}{_{--}} \sin^{3} \Theta _{B} \cos \Theta _{B}
 e^{
\imath \Phi _{R} }
\end{eqnarray}

The second part of the mixed helicity-properties contribution is
\begin{eqnarray}
I_{+-,++}^{g}= -  \frac{8p^2 m}{s^{3/2}} \,\, c(s,\Theta _{B})
 \,\, {\mathbf{R}}_{++}(F_{b}-\imath H_{b}) \sin^{3} \Theta _{B} \cos
\Theta _{B} e^{- \imath ( \Phi _{R} + \phi )}
\end{eqnarray}
\begin{eqnarray}
I_{-+,--}^{g}=  \frac{8p^2 m}{s^{3/2}} \,\,  c(s,\Theta _{B}) \,\,
{\mathbf{R}}_{--}(F_{b}+\imath H_{b}) \sin^{3} \Theta _{B} \cos
\Theta _{B} e^{ \imath ( \Phi _{R} + \phi )}
\end{eqnarray}
\begin{eqnarray}
I_{+-,--}^{g}=    \frac{8p^2 m}{s^{3/2}} \,\,  c(s,\Theta _{B})
\,\, {(F_{a}+\imath H_{a})}
 \overline{{{\mathbf{R}%
}}}{_{--}} \sin^{3} \Theta _{B} \cos \Theta _{B}
 e^{-
\imath \Phi _{R} }
\end{eqnarray}
\begin{eqnarray}
I_{-+,++}^{g}= -   \frac{8p^2 m}{s^{3/2}} \,\, c(s,\Theta _{B})
\,\, {(F_{a}-\imath H_{a})}
 \overline{{{\mathbf{R}%
}}}{_{++}} \sin^{3} \Theta _{B} \cos \Theta _{B}
 e^{
\imath \Phi _{R} }
\end{eqnarray}

The helicity-flip contribution is
\begin{eqnarray}
I_{++,++}^{g} =
\frac{4m^2}{s}  \,\, c(s,\Theta _{B})  \,\,  {\mathbf{%
R}_{++}\overline{\mathbf{R}}_{++}} [ 1 + \frac{4p^2}{s} ( 1+ \sin
^{4}\Theta _{B} ) ]  \\
I_{--,--}^{g} =
\frac{4m^2}{s}  \,\, c(s,\Theta _{B})  \,\, {\mathbf{%
R}_{--}\overline{\mathbf{R}}_{--}}
 [ 1 + \frac{4p^2}{s} ( 1+ \sin ^{4}\Theta _{B} ) ] \\
I_{++,--}^{g}  = \frac{4 m^2}{s} \,\, c(s,\Theta _{B}) \,\,
e^{\imath \phi } {\mathbf{r}_{+-}} \overline{{\mathbf{r}}}_{+-}
[ 1 - \frac{4p^2}{s} ( 1+ \sin ^{4}\Theta _{B} ) ]  \\
I_{--,++}^{g} = \frac{4 m^2}{s} \,\, c(s,\Theta _{B}) \,\,
e^{-\imath \phi } {\mathbf{r}_{-+}} \overline{{\mathbf{r}}}_{-+} [
1 - \frac{4p^2}{s} ( 1+ \sin ^{4}\Theta _{B} ) ]
\end{eqnarray}

\subsection{ Lepton-plus-Jets Channel:  $ \lambda_b = -1/2$,
$\lambda_{\bar{b}} = +1/2$
 \newline Dominance }

From the perspective of specific $t \bar{t}$ decay channels and/or
specific helicity amplitude tests, one can use the above results
to investigate various BR-S2SC functions.  For instance, in this
paper, we are most interested in the ``lepton-plus-jets channel"
and in tests for the relative sign of or for measurement of a
possible non-trivial phase between the $\lambda _{b}=-1/2$
helicity amplitudes for $t\rightarrow W^{+}b $. We assume
that the $\lambda _{b}=-1/2$ and $\lambda \overline{_{b}}$ $%
=1/2$ contributions dominate.

\subsubsection{$t_{1}\rightarrow W_{1}^{+}b\rightarrow (l^{+}\nu
)b$}

For the case $t_{1}\rightarrow W_{1}^{+}b\rightarrow (l^{+}\nu )b$, \ with $%
W_{2}^{-}$ decaying into hadronic jets, we separate the intensity
contributions into two parts: \  ``signal terms''
$\widetilde{I}|_{sig}$ which depend on $\Gamma
_{R}(0,-1)$ and $\Gamma _{I}(0,-1)$, and ``background terms'' $\widetilde{I|}%
_{0}$ which depend on $\Gamma (0,0)$ and $\Gamma (-1,-1)$.\ In
this section, we use a tilde accent
on $\widetilde{I|}%
_{0}, \ldots $  to denote the integration over the $\theta _{b}$, $%
{\phi }_{b}$ variables. By eqs(95-97) of ``I", this
integration directly projects out the various $ \overline{%
\Gamma } ( {\lambda_W}, {\lambda_W}^{'}) $ dependencies.

We find for the helicity-conserving contribution
\begin{eqnarray}
(\widetilde{I}_{+-,+-}^{g}+\widetilde{I}_{-+,-+}^{g})|_{0} &=&
\frac{8\pi p^2}{3s} \,\, c(s,\Theta _{B}) \,\,  \sin ^{2} \Theta
_{B} ( 1 + \cos ^{2} \Theta _{B}   )
\end{eqnarray}
\begin{eqnarray}
&&\left\{
\begin{array}{c}
\frac{1}{2}\Gamma (0,0)\sin ^{2}\theta _{a}[\overline{\Gamma
}(0,0)(1+\cos \theta _{1}^{t}\cos \theta
_{2}^{t})+\overline{\Gamma }(1,1)(1-\cos \theta
_{1}^{t}\cos \theta _{2}^{t})] \\
+\Gamma (-1,-1)\sin ^{4}\frac{\theta _{a}}{2}[\overline{\Gamma
}(0,0)(1-\cos \theta _{1}^{t}\cos \theta
_{2}^{t})+\overline{\Gamma }(1,1)(1+\cos \theta _{1}^{t}\cos
\theta _{2}^{t})] \nonumber
\end{array}
\right\}
\end{eqnarray}
\begin{eqnarray}
(\widetilde{I}_{+-,+-}^{g}+\widetilde{I}_{-+,-+}^{g})|_{sig} &=&
\frac{8\sqrt{2}\pi p^2}{3s} \,\, c(s,\Theta _{B}) \,\, \sin ^{2}
\Theta _{B} ( 1 + \cos ^{2} \Theta _{B}   ) \sin \theta
_{1}^{t}\cos \theta _{2}^{t}\sin
\theta _{a}\sin ^{2}\frac{\theta _{a}}{2} \nonumber \\
&&\left\{ -\Gamma _{R}(0,-1)\cos {\phi }_{a}+\Gamma _{I}(0,-1)\sin
{\phi }_{a}\right\} [\overline{\Gamma }(0,0)-\overline{\Gamma }%
(1,1)]
\end{eqnarray}
\begin{eqnarray}
(\widetilde{I}_{+-,-+}^{g}+\widetilde{I}_{-+,+-}^{g})|_{0} &=& -
\frac{8\pi p^2}{3 s} \,\, c(s,\Theta _{B}) \,\,  \sin ^{4} \Theta
_{B} \cos (2\Phi _{R}+\phi )\sin \theta _{1}^{t}\sin \theta
_{2}^{t} \\
&&\left\{ \frac{1}{2}\Gamma (0,0)\sin ^{2}\theta _{a}-\Gamma (-1,-1)\sin ^{4}%
\frac{\theta _{a}}{2}\right\} [\overline{\Gamma }(0,0)-\overline{\Gamma }%
(1,1)]\nonumber
\end{eqnarray}
\begin{eqnarray}
(\widetilde{I}_{+-,-+}^{g}+\widetilde{I}_{-+,+-}^{g})|_{sig} &=&-
\frac{8\sqrt{2} \pi p^2}{3 s} \,\, c(s,\Theta _{B}) \,\,  \sin
^{4} \Theta _{B}
\sin \theta _{2}^{t}\sin \theta _{a}\sin ^{2}\frac{%
\theta _{a}}{2}
\end{eqnarray}
\begin{eqnarray}
&&\left\{
\begin{array}{c}
\cos (2\Phi _{R}+\phi )\cos \theta _{1}^{t}\left\{ \Gamma
_{R}(0,-1)\cos {\phi }_{a}-\Gamma _{I}(0,-1)\sin {\phi
}_{a}\right\}
\\
+\sin (2\Phi _{R}+\phi )\left\{ \Gamma _{R}(0,-1)\sin {\phi }%
_{a}+\Gamma _{I}(0,-1)\cos {\phi }_{a}\right\}
\end{array}
\right\} [ \overline{\Gamma }(0,0)-\overline{\Gamma }(1,1)]
\nonumber
\end{eqnarray}

We collect the mixed-helicity contributions in real sums:
\begin{eqnarray}
(\widetilde{I}_{-+,++}^{g}+\widetilde{I}_{--,+-}^{g}+
\widetilde{I}_{+-,--}^{g}+\widetilde{I}_{++,-+}^{g})|_{0} &=&
\frac{32\pi p^2 m}{3 s^{3/2}} \,\,  c(s,\Theta _{B}) \,\, \sin
^{3} \Theta _{B}   \cos \Theta _{B}
\end{eqnarray}
\begin{eqnarray}
\cos \Phi _{R}\sin \theta _{1}^{t}\cos \theta _{2}^{t} \left\{
\frac{1}{2}\Gamma (0,0)\sin ^{2}\theta _{a}-\Gamma (-1,-1)\sin ^{4}%
\frac{\theta _{a}}{2}\right\} [\overline{\Gamma }(0,0)-\overline{\Gamma }%
(1,1)]\nonumber
\end{eqnarray}
\begin{eqnarray}
(\widetilde{I}_{-+,++}^{g}+\widetilde{I}_{--,+-}^{g}+
\widetilde{I}_{+-,--}^{g}+\widetilde{I}_{++,-+}^{g})|_{sig} &=&
\frac{32\sqrt{2} \pi p^2 m}{3 s^{3/2}} \,\, c(s,\Theta _{B}) \,\,
\sin ^{3} \Theta _{B} \cos  \Theta _{B}
\end{eqnarray}
\begin{eqnarray}
\cos \theta _{2}^{t}\sin \theta _{a}\sin ^{2}\frac{\theta _{a}}{2}
\left\{
\begin{array}{c}
\cos \theta _{1}^{t}\left\{ \Gamma _{R}(0,-1)\cos {\phi }%
_{a}-\Gamma _{I}(0,-1)\sin {\phi }_{a}\right\} \cos \Phi _{R} \\
+\left\{ \Gamma _{R}(0,-1)\sin {\phi }_{a}+\Gamma _{I}(0,-1)\cos
{\phi }_{a}\right\} \sin \Phi _{R}
\end{array}
\right\} [\overline{\Gamma }(0,0)-\overline{\Gamma
}(1,1)]\nonumber
\end{eqnarray}
\begin{eqnarray}
(\widetilde{I}_{++,+-}^{g}+\widetilde{I}_{--,-+}^{g}+
\widetilde{I}_{+-,++}^{g}+\widetilde{I}_{-+,--}^{g})|_{0} &=& -
\frac{32\pi p^2 m}{3 s^{3/2}} \,\,  c(s,\Theta _{B}) \,\,   \sin
^{3} \Theta _{B}   \cos \Theta _{B}
\end{eqnarray}
\begin{eqnarray}
\cos (\Phi _{R}+\phi )\cos \theta _{1}^{t}\sin \theta _{2}^{t}
\left\{ \frac{1}{2}\Gamma (0,0)\sin ^{2}\theta _{a}-\Gamma (-1,-1)\sin ^{4}%
\frac{\theta _{a}}{2}\right\} [\overline{\Gamma }(0,0)-\overline{\Gamma }%
(1,1)]\nonumber
\end{eqnarray}
\begin{eqnarray}
(\widetilde{I}_{++,+-}^{g}+\widetilde{I}_{--,-+}^{g}+
\widetilde{I}_{+-,++}^{g}+\widetilde{I}_{-+,--}^{g})|_{sig} &=&
\frac{32\sqrt{2} \pi p^2 m}{3 s^{3/2}} \,\,  c(s,\Theta _{B}) \,\,
\sin ^{3} \Theta _{B} \cos  \Theta _{B}
\end{eqnarray}
\begin{eqnarray}
\cos (\Phi _{R}+\phi )\sin \theta _{1}^{t}\sin \theta _{2}^{t}\sin
\theta _{a}\sin ^{2}\frac{\theta _{a}}{2} \left\{ \Gamma
_{R}(0,-1)\cos {\phi }_{a}-\Gamma _{I}(0,-1)\sin
{\phi }_{a}\right\} [\overline{\Gamma }(0,0)-\overline{\Gamma }%
(1,1)]\nonumber
\end{eqnarray}

The helicity-flip contributions are
\begin{eqnarray}
(\widetilde{I}_{++,++}^{g}+\widetilde{I}_{--,--}^{g})|_{0} &=&
\frac{8\pi m^2}{3 s} \,\, c(s,\Theta _{B}) \,\,  [ 1+
\frac{4p^2}{s} ( 1 + \sin ^{4} \Theta _{B}   ) ]
\end{eqnarray}
\begin{eqnarray}
&&\left\{
\begin{array}{c}
\frac{1}{2}\Gamma (0,0)\sin ^{2}\theta _{a}[\overline{\Gamma
}(0,0)(1-\cos \theta _{1}^{t}\cos \theta
_{2}^{t})+\overline{\Gamma }(1,1)(1+\cos \theta
_{1}^{t}\cos \theta _{2}^{t})] \\
+\Gamma (-1,-1)\sin ^{4}\frac{\theta _{a}}{2}[\overline{\Gamma
}(0,0)(1+\cos \theta _{1}^{t}\cos \theta
_{2}^{t})+\overline{\Gamma }(1,1)(1-\cos \theta _{1}^{t}\cos
\theta _{2}^{t})]  \nonumber
\end{array}
\right\}
\end{eqnarray}
\begin{eqnarray}
(\widetilde{I}_{++,++}^{g}+\widetilde{I}_{--,--}^{g})|_{sig} &=&
\frac{8\sqrt{2}\pi m^2}{3 s} \,\,  c(s,\Theta _{B}) \,\,  [ 1+
\frac{4p^2}{s} ( 1 + \sin ^{4} \Theta _{B}   ) ]
\end{eqnarray}
\begin{eqnarray}
\sin \theta _{1}^{t}\cos \theta _{2}^{t}\sin \theta _{a}\sin
^{2}\frac{\theta _{a}}{2} \left\{ \Gamma _{R}(0,-1)\cos {\phi
}_{a}-\Gamma _{I}(0,-1)\sin
{\phi }_{a}\right\} [\overline{\Gamma }(0,0)-\overline{\Gamma }%
(1,1)]\nonumber
\end{eqnarray}
\begin{eqnarray}
(\widetilde{I}_{++,--}^{g}+\widetilde{I}_{--,++}^{g})|_{0} &=&
\frac{8\pi m^2}{3 s} \,\, c(s,\Theta _{B}) \,\,  [ 1-
\frac{4p^2}{s} ( 1 + \sin ^{4} \Theta _{B}   ) ]
\end{eqnarray}
\begin{eqnarray}
\cos \phi \sin \theta _{1}^{t}\sin \theta _{2}^{t} \left\{
-\frac{1}{2}\Gamma (0,0)\sin ^{2}\theta _{a}+\Gamma (-1,-1)\sin
^{4}\frac{\theta _{a}}{2}\right\} [\overline{\Gamma }(0,0)-\overline{\Gamma }%
(1,1)]\nonumber
\end{eqnarray}
\begin{eqnarray}
(\widetilde{I}_{++,--}^{g}+\widetilde{I}_{--,++}^{g})|_{sig} &=&
\frac{8\sqrt{2}\pi m^2}{3 s} \,\, c(s,\Theta _{B}) \,\,  [ 1-
\frac{4p^2}{s} ( 1 + \sin ^{4} \Theta _{B}   ) ] \sin \theta
_{2}^{t}\sin \theta _{a}
\end{eqnarray}
\begin{eqnarray}
\sin ^{2}\frac{\theta _{a}}{2} \left\{
\begin{array}{c}
\cos \phi \cos \theta _{1}^{t}\left\{ -\Gamma _{R}(0,-1)\cos {\phi
}_{a}+\Gamma _{I}(0,-1)\sin {\phi }_{a}\right\}  \\
+\sin \phi \left\{ \Gamma _{R}(0,-1)\sin {\phi }_{a}+\Gamma
_{I}(0,-1)\cos {\phi }_{a}\right\}
\end{array}
\right\} [\overline{\Gamma }(0,0)-\overline{\Gamma
}(1,1)]\nonumber
\end{eqnarray}

\subsection{$\overline{t}_{2}\rightarrow W_{2}^{-}%
\overline{b}\rightarrow (l^{-}\bar{\nu} )\overline{b}$}

For the\ $CP$-conjugate process $\overline{t}_{2}\rightarrow W_{2}^{-}%
\overline{b}\rightarrow (l^{-}\bar{\nu} )\overline{b}$, \ with
$W_{1}^{+}$ decaying into hadronic jets, we similarly separate the
contributions: \
``signal terms'' $\widetilde{\overline{I}|}_{sig}$depending on $\overline{%
\Gamma }_{R}(0,1)$ and $\overline{\Gamma }_{I}(0,1)$, and
``background
terms'' $\widetilde{\overline{I}}|_{0}$ depending on $\overline{\Gamma }%
(0,0) $ and $\overline{\Gamma }(1,1)$.

By integration over $\theta _{a}$, ${\phi }_{a}$, we find for the
helicity-conserving contribution
\begin{eqnarray}
(\widetilde{\overline{I}}_{+-,+-}^{g}+\widetilde{\overline{I}}_{-+,-+}^{g})|_{0}
&=& \frac{8\pi p^2}{3 s}  \,\, c(s,\Theta _{B}) \,\, \sin ^{2}
\Theta _{B} ( 1 + \cos ^{2} \Theta _{B}   )
\end{eqnarray}
\begin{eqnarray}
&&\left\{
\begin{array}{c}
\frac{1}{2}\overline{\Gamma }(0,0)\sin ^{2}\theta _{b}[\Gamma
(0,0)(1+\cos \theta _{1}^{t}\cos \theta _{2}^{t})+\Gamma
(-1,-1)(1-\cos \theta
_{1}^{t}\cos \theta _{2}^{t})] \\
+\overline{\Gamma }(1,1)\sin ^{4}\frac{\theta _{b}}{2}[\Gamma
(0,0)(1-\cos \theta _{1}^{t}\cos \theta _{2}^{t})+\Gamma
(-1,-1)(1+\cos \theta _{1}^{t}\cos \theta _{2}^{t})]
\end{array}
\right\} \nonumber
\end{eqnarray}
\begin{eqnarray}
(\widetilde{\overline{I}}_{+-,+-}^{g}+\widetilde{\overline{I}}_{-+,-+}^{g})|_{sig} &=&%
- \frac{8\sqrt{2}\pi p^2}{3s} \,\, c(s,\Theta _{B}) \,\,  \sin
^{2} \Theta _{B} ( 1 + \cos ^{2} \Theta _{B}   ) \cos \theta
_{1}^{t}\sin \theta _{2}^{t}\sin \theta _{b}\sin ^{2}\frac{\theta _{b}}{2} \nonumber \\
&&\left\{ \overline{\Gamma }_{R}(0,1)\cos {\phi }_{b}+ \overline{%
\Gamma }_{I}(0,1)\sin {\phi }_{b}\right\} [\Gamma (0,0)-\Gamma
(-1,-1)]
\end{eqnarray}
\begin{eqnarray}
(\widetilde{\overline{I}}_{+-,-+}^{g}+\widetilde{\overline{I}}_{-+,+-}^{g})|_{0}
&=& - \frac{8\pi p^2}{3 s} \,\, c(s,\Theta _{B}) \,\, \sin ^{4}
\Theta _{B} \cos (2\Phi _{R}+\phi )\sin \theta _{1}^{t}\sin \theta
_{2}^{t} \\
&&\left\{ \frac{1}{2}\overline{\Gamma }(0,0)\sin ^{2}\theta _{b}-\overline{%
\Gamma }(1,1)\sin ^{4}\frac{\theta _{b}}{2}\right\} [\Gamma
(0,0)-\Gamma (-1,-1)]\nonumber
\end{eqnarray}
\begin{eqnarray}
(\widetilde{\overline{I}}_{+-,-+}^{g}+\widetilde{\overline{I}}_{-+,+-}^{g})|_{sig}
&=& -%
 \frac{8\sqrt{2}\pi p^2}{3 s} \,\,  c(s,\Theta _{B}) \,\,  \sin
^{4} \Theta _{B} \sin \theta _{1}^{t}\sin \theta _{b}\sin
^{2}\frac{\theta _{b}}{2}
\end{eqnarray}
\begin{eqnarray}
&&\left\{
\begin{array}{c}
\cos (2\Phi _{R}+\phi )\cos \theta _{2}^{t}\left\{ \overline{\Gamma }%
_{R}(0,1)\cos {\phi }_{b}+\overline{\Gamma }_{I}(0,1)\sin
{\phi }_{b}\right\}  \\
-\sin (2\Phi _{R}+\phi )\left\{ \overline{\Gamma }_{R}(0,1)\sin {%
\phi }_{b}-\overline{\Gamma }_{I}(0,1)\cos {\phi }_{b}\right\}
\end{array}
\right\}  [\Gamma (0,0)-\Gamma (-1,-1)] \nonumber
\end{eqnarray}

The mixed-helicity contributions are
\begin{eqnarray}
(\widetilde{\overline{I}}_{-+,++}^{g}+\widetilde{\overline{I}}_{--,+-}^{g}+
\widetilde{\overline{I}}_{+-,--}^{g}+\widetilde{\overline{I}}_{++,-+}^{g})|_{0}
 &=&  \frac{32\pi p^2 m}{3 s^{3/2}} \,\,  c(E,\Theta _{B}) \,\,  \sin ^{3}
\Theta _{B}   \cos  \Theta _{B}
\end{eqnarray}
\begin{eqnarray}
\cos \Phi _{R} \sin \theta _{1}^{t}\cos \theta _{2}^{t}
\left\{ \frac{1}{2}\overline{\Gamma }(0,0)\sin ^{2}\theta _{b}-\overline{%
\Gamma }(1,1)\sin ^{4}\frac{\theta _{b}}{2}\right\} [{\Gamma }(0,0)-%
{\Gamma }(-1,-1)]\nonumber
\end{eqnarray}
\begin{eqnarray}
(\widetilde{\overline{I}}_{-+,++}^{g}+\widetilde{\overline{I}}_{--,+-}^{g}+
\widetilde{\overline{I}}_{+-,--}^{g}+\widetilde{\overline{I}}_{++,-+}^{g})|_{sig}
&=& - \frac{32\sqrt{2} \pi p^2 m}{3 s^{3/2}} \,\, c(s,\Theta _{B})
\,\, \sin ^{3} \Theta _{B}   \cos  \Theta _{B}
\end{eqnarray}
\begin{eqnarray}
\cos \Phi _{R}\sin \theta _{1}^{t}\sin \theta _{2}^{t}\sin \theta
_{b}\sin ^{2}\frac{\theta _{b}}{2}
\left\{ \overline{\Gamma }_{R}(0,1)\cos {\phi }_{b}+\overline{%
\Gamma }_{I}(0,1)\sin {\phi }_{b}\right\} [\Gamma (0,0)-\Gamma
(-1,-1)]\nonumber
\end{eqnarray}
\begin{eqnarray}
(\widetilde{\overline{I}}_{++,+-}^{g}+\widetilde{\overline{I}}_{--,-+}^{g}+
\widetilde{\overline{I}}_{+-,++}^{g}+\widetilde{\overline{I}}_{-+,--}^{g})|_{0}
 &=& - \frac{32\pi p^2 m}{3 s^{3/2}} \,\, c(s,\Theta _{B}) \,\, \sin ^{3}
\Theta _{B}   \cos  \Theta _{B}
\end{eqnarray}
\begin{eqnarray}
\cos (\Phi _{R}+\phi )\cos \theta _{1}^{t}\sin \theta _{2}^{t}
\left\{ \frac{1}{2}\overline{\Gamma }(0,0)\sin ^{2}\theta _{b}-\overline{%
\Gamma }(1,1)\sin ^{4}\frac{\theta _{b}}{2}\right\} [\Gamma
(0,0)-\Gamma (-1,-1)]\nonumber
\end{eqnarray}
\begin{eqnarray}
(\widetilde{\overline{I}}_{++,+-}^{g}+\widetilde{\overline{I}}_{--,-+}^{g}+
\widetilde{\overline{I}}_{+-,++}^{g}+\widetilde{\overline{I}}_{-+,--}^{g})|_{sig}
 = - \frac{32\sqrt{2}\pi p^2 m}{3 s^{3/2}} \,\, c(s,\Theta _{B})  \sin ^{3}
\Theta _{B}   \cos  \Theta _{B} \cos \theta _{1}^{t}
\end{eqnarray}
\begin{eqnarray}
\sin \theta _{b}\sin ^{2}\frac{%
\theta _{b}}{2}
\left\{
\begin{array}{c}
\cos \theta _{2}^{t}\left\{ \overline{\Gamma }_{R}(0,1)\cos {\phi }%
_{b}+\overline{\Gamma }_{I}(0,1)\sin {\phi }_{b}\right\} \cos
(\Phi _{R}+\phi ) \\
+\left\{ -\overline{\Gamma }_{R}(0,1)\sin {\phi }_{b}+\overline{%
\Gamma }_{I}(0,1)\cos {\phi }_{b}\right\} \sin (\Phi _{R}+\phi )
\end{array}
\right\} [\Gamma (0,0)-\Gamma (-1,-1)]\nonumber
\end{eqnarray}

The helicity-flip contributions are
\begin{eqnarray}
(\widetilde{\overline{I}}_{++,++}^{g}+\widetilde{\overline{I}}_{--,--}^{g})|_{0}
&=& \frac{8\pi m^2}{3 s} \,\, c(s,\Theta _{B}) \,\, [ 1+ \frac{4
p^2}{s} ( 1 + \sin ^{4} \Theta _{B}   ) ]
\end{eqnarray}
\begin{eqnarray}
&&\left\{
\begin{array}{c}
\frac{1}{2}\overline{\Gamma }(0,0)\sin ^{2}\theta _{b}[\Gamma
(0,0)(1-\cos \theta _{1}^{t}\cos \theta _{2}^{t})+\Gamma
(-1,-1)(1+\cos \theta
_{1}^{t}\cos \theta _{2}^{t})] \\
+\overline{\Gamma }(1,1)\sin ^{4} \frac{\theta _{b}}{2} [\Gamma
(0,0)(1+\cos \theta _{1}^{t}\cos \theta _{2}^{t})+\Gamma
(-1,-1)(1-\cos \theta _{1}^{t}\cos \theta _{2}^{t})]
\end{array}
\right\} \nonumber
\end{eqnarray}
\begin{eqnarray}
(\widetilde{\overline{I}}_{++,++}^{g}+\widetilde{\overline{I}}_{--,--}^{g})|_{sig}
&=& \frac{8\sqrt{2}\pi m^2}{3 s} \,\, c(s,\Theta _{B}) \,\, [ 1+
\frac{4 p^2}{s} ( 1 + \sin ^{4} \Theta _{B}   ) ]
\end{eqnarray}
\begin{eqnarray}
\cos \theta _{1}^{t}\sin \theta _{2}^{t}\sin \theta _{b}\sin
^{2}\frac{\theta _{b}}{2}
\left\{ \overline{\Gamma }_{R}(0,1)\cos {\phi }_{b}+\overline{%
\Gamma }_{I}(0,1)\sin {\phi }_{b}\right\} [\Gamma (0,0)-\Gamma
(-1,-1)]  \nonumber
\end{eqnarray}
\begin{eqnarray}
(\widetilde{\overline{I}}_{++,--}^{g}+\widetilde{\overline{I}}_{--,++}^{g})|_{0}
&=&  \frac{8\pi m^2}{3 s} \,\,  c(s,\Theta _{B}) \,\,  [ 1 -
\frac{4 p^2}{s} ( 1 + \sin ^{4} \Theta _{B}   ) ]
\end{eqnarray}
\begin{eqnarray}
\cos \phi \sin \theta _{1}^{t}\sin \theta _{2}^{t}
\left\{ -\frac{1}{2}\overline{\Gamma }(0,0)\sin ^{2}\theta _{b}+\overline{%
\Gamma }(1,1)\sin ^{4}\frac{\theta _{b}}{2}\right\} [\Gamma
(0,0)-\Gamma (-1,-1)]\nonumber
\end{eqnarray}
\begin{eqnarray}
(\widetilde{\overline{I}}_{++,--}^{g}+\widetilde{\overline{I}}_{--,++}^{g})|_{sig}
&=& - \frac{8\sqrt{2} \pi m^2}{3 s} \,\, c(s,\Theta _{B}) \,\,  [
1 - \frac{4 p^2}{s} ( 1 + \sin ^{4} \Theta _{B}   ) ] \sin \theta
_{1}^{t}\sin \theta _{b}
\end{eqnarray}
\begin{eqnarray}
\sin ^{2}\frac{\theta _{b}%
}{2}
\left\{
\begin{array}{c}
\cos \phi \cos \theta _{2}^{t}\left\{ \overline{\Gamma
}_{R}(0,1)\cos
{\phi }_{b}+\overline{\Gamma }_{I}(0,1)\sin {\phi }%
_{b}\right\}  \\
-\sin \phi \left\{ \overline{\Gamma }_{R}(0,1)\sin {\phi }_{b}-%
\overline{\Gamma }_{I}(0,1)\cos {\phi }_{b}\right\}
\end{array}
\right\} [\Gamma (0,0)-\Gamma (-1,-1)]\nonumber
\end{eqnarray}

\subsection{$\Gamma( \lambda_W, {\lambda_W}^{'} )$ tests versus
angular dependence}

In summary, with beam-referencing, for the $t_{1}\rightarrow
W_{1}^{+}b\rightarrow (l^{+}{\nu })b$ case there are six
``background terms'' depending on  $\Gamma (0,0)$ and $\Gamma
(-1,-1)$, and also six ``signal terms'' depending on $\Gamma
_{R,I}(0,-1)$.  As a consequence of Lorentz invariance, there are
associated kinematic factors with simple angular dependence which
can be used to isolate and measure these four $\Gamma ^{\prime
}$s. The following patterns, (i) and (ii), for these BR-S2SC
contributions $\widetilde{I} _{\lambda _1\lambda _2;\lambda
_1^{^{\prime }}\lambda _2^{^{\prime }}}^{g,q}$ occur identically
in gluon-production and in quark-production:

(i) $\theta _{a}$ polar-angle dependence:

The coefficients of  $%
\Gamma (0,0) \Big / \Gamma (-1,-1)\Big / \Gamma _{R,I}(0,-1)$ vary
relatively as the $W$-decay \newline $d_{m m^{\prime}} ^1 (
\theta_a )$-squared-intensity-ratios
\begin{eqnarray}
\frac{1}{2}\sin ^{2}\theta _{a}\Big / \left[ \sin ^{4}\frac{\theta _{a}}{2}%
\right] \Big / \left\{ \frac{1}{\sqrt{2}}\sin \theta _{a}\sin ^{2}\frac{%
\theta _{a}}{2}\right\} = \nonumber \\
2(1+\cos \theta _{a}) \Big / \left[ 1-\cos \theta
_{a}\right] \Big / \left\{ \sqrt{2(1+\cos \theta _{a})(1-\cos \theta _{a})}%
=\sqrt{2}\sin \theta _{a}\right\}
\end{eqnarray}

(ii) $\phi _{a}$ azimuthal-angle dependence in the ``signal
terms'' :

The coefficients of $\Gamma _{R}(0,-1) \Big / \Gamma _{I}(0,-1)$ \
vary as
\begin{equation}
\cos \phi _{a} \Big / \sin \phi _{a}
\end{equation}
in each of the signal terms, i.e. as in the four-angle
gluon-signal term (3) there is a factor $\left\{ \Gamma
_{R}(0,-1)\cos {\phi }_{a}-\Gamma _{I}(0,-1)\sin {\phi
}_{a}\right\}$. However, in three terms there are also $ \Gamma
_{R,I}(0,-1)$'s with the opposite association of these $\cos \phi
_{a}$, $ \sin \phi _{a}$ factors, i.e. in a factor
\newline
$\left\{ \Gamma _{R}(0,-1)\sin {\phi }_{a}+\Gamma
_{I}(0,-1)\cos {\phi }_{a}\right\}$. This opposite association
occurs in half the ``signal" contributions: the helicity
conserving
$(\widetilde{I}_{+-,-+}^{g}+\widetilde{I}_{-+,+-}^{g})|_{sig}$ of
(54), the mixed-helicity
$(\widetilde{I}_{-+,++}^{g}+\widetilde{I}_{--,+-}^{g}+
\widetilde{I}_{+-,--}^{g}+\widetilde{I}_{++,-+}^{g})|_{sig}$ of
(56), and in the helicity-flip $(\widetilde{I}_{++,--}^{g}
+\widetilde{I}_{--,++}^{g})|_{sig}$ of (62), along with a different $%
\Phi _{R}$ and/or $\phi $ dependence.  This different angular
dependence might be used in an empirical separation of these terms
from the terms with the normal $\phi _{a} $ association.  This
different $\Phi _{R}$ and/or $\phi $ dependence is the reason that
only the $\left\{ \Gamma _{R}(0,-1)\cos {\phi }_{a}-\Gamma
_{I}(0,-1)\sin {\phi }_{a}\right\}$ factor appears in the signal
terms in the four-angle and five-angle BR-S2SC functions.

There are analogous CP-conjugate patterns in $\theta_{b}$ and
$\phi_{b} $ for the CP-conjugate case $\bar{t_{2}} \rightarrow
W_{2}^{-}\bar{b} \rightarrow (l^{-}{\bar{\nu} })\bar{b}$. There is
only the factor $ \left\{ \overline{\Gamma }_{R}(0,1)\cos \phi
_{b}+\overline{\Gamma }_{I}(0,1)\sin \phi _{b}\right\}$ in the
four-angle and five-angle BR-S2SC functions. In the CP-conjugate
case the patterns are also identical in gluon-production and in
quark-production.

\subsubsection{Six-angle $\widetilde{\mathcal{H}}_{i}^{g}\quad
$distributions:  $\phi $ dependence}

To reduce the number of angles to obtain six-angle BR-S2SC
functions, we integrate out either the beam-referencing
azimuthal-angle $\Phi _{R}$ or the angle $\phi $ between the $t_1$
and $\bar{t_2}$ decay planes .  We only list the non-vanishing
contributions. To clearly label the successive contributions, we
again use the index $i=(\lambda _{1}\lambda _{2},\lambda
_{1}^{^{\prime }}\lambda _{2}^{^{\prime }})$.

We first consider the $t\longrightarrow W^{+}b\longrightarrow
l^{+}\nu b$ channel with the $W^{-}$ decaying hadronically.  For
\begin{eqnarray}
\widetilde{\mathcal{H}}_{i}^{g} \equiv \int_{0}^{2\pi }d\Phi
_{R}\widetilde{I}_{i}^{g}
\end{eqnarray}
the non-vanishing six-angle contributions are proportional to the
``background and signal parts'' of the above expressions with
\begin{equation}
\widetilde{\mathcal{H}}_{i}^{g}=2\pi \widetilde{I}_{i}^{g}
\end{equation}
For the helicity-conserving, $i=(+-,+-)+(-+,-+)$ with no $\phi $
dependence.  For the helicity-flip $i=(++,++)+(--,--)$ with no
$\phi $ dependence; and $i=(++,--)+(--,++)$ with a $\cos\phi $
dependence in the background part, and both $\cos\phi $ and
$\sin\phi $ dependence in the signal part. The mixed-helicity
contributions vanish.

\subsubsection{Six-angle $(\widetilde{\mathcal{H}}_{i}^{g})^{^{\prime }}\quad $%
distributions:  $\Phi_R$ dependence}

If instead the $\phi $ dependence is integrated out
\begin{eqnarray}
(\widetilde{\mathcal{H}}_{i}^{g})^{^{\prime }}\equiv
\int_{0}^{2\pi }d\phi \widetilde{I}_{i}^{g}
\end{eqnarray}
the non-vanishing six-angle contributions are also proportional to
the above expressions:
\begin{equation}
(\widetilde{\mathcal{H}}_{i}^{g})^{^{\prime }}=2\pi
\widetilde{I}_{i}^{g}
\end{equation}
For the helicity-conserving, $i=(+-,+-)+(-+,-+)$
with no $\Phi _{R}$ dependence.  For the mixed-helicity, $%
i=(-+,++)+(--,+-)+(+-,--)+(++,-+)$  with a $ \cos \Phi _{R}$
dependence in the background part, and both $ \cos \Phi _{R}$ and
$ \sin \Phi _{R}$ dependence in
the signal part. For the helicity-flip $i=(++,++)+(--,--)$ with no  $%
\Phi _{R}$ dependence.

\subsubsection{Five-angle $\widetilde{\mathcal{G}}_{i}^{g} $ distributions:}

If both the $\Phi _{R}$ and $\phi $ dependence are integrated out,
the five-angle distribution \newline
$ \{\Theta _{B,}\theta
_{1}^{t},\theta _{2}^{t},\theta _{a},\phi _{a}\}$ is
\begin{equation}
\widetilde{\mathcal{G}}_{i}^{g}\equiv \int_{0}^{2\pi }d\phi
\int_{0}^{2\pi }d\Phi _{R}\widetilde{I}_{i}^{g}
\end{equation}
The terms in these expressions only arise from the
helicity-conserving  (51,52), and from the helicity-flip (59, 60):
\begin{equation}
\widetilde{\mathcal{G}}^{g}|_{0}=4\pi ^{2}[ (\widetilde{I}_{+-,+-}^{g}+%
\widetilde{I}_{-+,-+}^{g})|_{0}+(\widetilde{I}_{++,++}^{g}+\widetilde{I}%
_{--,--}^{g})|_{0} ]
\end{equation}
\begin{equation}
\widetilde{\mathcal{G}}^{g}|_{sig}=4\pi ^{2}[ (\widetilde{I}%
_{+-,+-}^{g}+\widetilde{I}_{-+,-+}^{g})|_{sig}+(\widetilde{I}_{++,++}^{g}+%
\widetilde{I}%
_{--,--}^{g})|_{sig} ]
\end{equation}

We obtain
\begin{eqnarray}
\widetilde{\mathcal{G}}^{g}|_{0} &=&{}\frac{8\pi ^{3}}{3} \,\,
c(s,\Theta _{B}) \,\, \left\{ \frac{1}{2}\Gamma (0,0)\sin
^{2}\theta _{a}+\Gamma
(-1,-1)\sin ^{4}\frac{\theta _{a}}{2}\right\} [\overline{\Gamma }(0,0)+%
\overline{\Gamma }(1,1)]  \nonumber \\
&&\left\{ \widetilde{g}_{1}^{g}(s,\Theta _{B})+{\mathcal{R\,\,}}\widetilde{g}%
_{2}^{g}(s,\Theta _{B})\cos \theta _{1}^{t}\cos \theta
_{2}^{t}\right\}
\end{eqnarray}
\begin{eqnarray}
\widetilde{\mathcal{G}}^{g}|_{sig} &=&-{}\frac{8\sqrt{2}\pi ^{3}}{3}%
\,\, c(s,\Theta _{B}) \,\, \sin \theta _{a}\sin ^{2}\frac{\theta
_{a}}{2}\left\{
\Gamma _{R}(0,-1)\cos \phi _{a}-\Gamma _{I}(0,-1)\sin \phi _{a}\right\}  \nonumber \\
&&\lbrack \overline{\Gamma }(0,0)+\overline{\Gamma }(1,1)]\,{\mathcal{R\,\,}}%
\widetilde{g}_{2}^{g}(s,\Theta _{B})\sin \theta _{1}^{t}\cos
\theta _{2}^{t}
\end{eqnarray}
where the two gluon-beam-referencing factors $
\widetilde{g}_{1,2}^{g}(s,\Theta _{B}) $ are listed in (4, 5).

Note that versus the four-angle distributions listed in the
introduction, in this five-angle distribution
$\widetilde{\mathcal{G}}^{g} $ the gluon-beam-referencing factor $
\widetilde{g}_{2}^{g}(s,\Theta _{B}) $ appears in both the
background and the signal contributions.  For the quark-production
contribution to the five-angle distribution
$\widetilde{\mathcal{G}}^{q} $ of (111,112), the analogous
situation occurs in the quark-beam-referencing factors
$\widetilde{g}_{1,2}^{q}(s,\Theta _{B})$  versus the four-angle
distribution (9, 10). Likewise for the respective CP-conjugate
spin-correlation functions.

\subsubsection{Four-angle ${\mathcal{G}}_{i}^{g}$distributions:}

If the $\cos \theta _{1}^{t}$ dependence is integrated out, there
is a four-angle distribution in \newline $\{\Theta _{B},\theta
_{2}^{t},\theta _{a},\phi _{a}\} $ with the same ``$i$" values as
in (82,83),
\begin{eqnarray}
{\mathcal{G}}_{i}^{g} \equiv \int_{-1}^{1} d ( \cos \theta _
{1}^{t} ) \int_{0}^{2\pi }d\phi \int_{0}^{2\pi }d\Phi
_{R}   \widetilde{I}_{i}^{g} \\
= \int_{-1}^{1} d ( \cos \theta _{1}^{t} )
\widetilde{\mathcal{G}}_{i}^{g} \nonumber
\end{eqnarray}
 which is listed in the
introduction section in (2,3).

\subsubsection{Integrated BR-S2SC Distributions for $CP$ conjugate cases:}

For the $CP$-conjugate case in terms of $\{\Phi_{R},\phi,\Theta
_{B},\theta _{2}^{t}$, $\theta _{1}^{t}$, $\theta _{b}$, $\phi
_{b} \}$, the successively fewer angle distributions similarly
follow.  The non-vanishing terms are proportional to the preceding
``background and signal'' expressions (63-74). The six-angle
expression is
\begin{eqnarray}
\widetilde{\overline{\mathcal{H}}_{i}^{g}} \equiv \int_{0}^{2\pi
}d\Phi _{R} \widetilde{\overline{I}}_{i}^{g} &=& 2\pi
\widetilde{\overline{I}}_{i}^{g}
\end{eqnarray}
For the helicity-conserving, $i=(+-,+-)+(-+,-+)$ with no $\phi $
dependence.  For the helicity-flip, $i=(++,++)+(--,--)$ with no
$\phi $ dependence; and $i=(++,--)+(--,++)$ with a $\cos\phi $
dependence in the background part, and both $\cos\phi $ and
$\sin\phi $ dependence in the signal part.

If instead the $\phi $ dependence is integrated out, there is the
six-angle expression
\begin{eqnarray}
(\widetilde{\overline{\mathcal{H}}_{i}^{g}})^{^{\prime }}\equiv
\int_{0}^{2\pi }d\phi \widetilde{\overline{I}}_{i}^{g} &=& 2\pi
\widetilde{\overline{I}}_{i}^{g}
\end{eqnarray}
For the helicity-conserving, $i=(+-,+-)+(-+,-+)$
with no $\Phi _{R}$ dependence.  For the mixed-helicity, $%
i=(-+,++)+(--,+-)+(+-,--)+(++,-+)$  with a $ \cos \Phi _{R}$
dependence in the background part, and both $ \cos \Phi _{R}$ and
$ \sin \Phi _{R}$ dependence in
the signal part. For the helicity-flip, $i=(++,++)+(--,--)$ with no  $%
\Phi _{R}$ dependence.

If both the $\Phi _{R}$ and $\phi $ dependence are integrated out,
the five-angle distribution \newline $ \{\Theta _{B,}\theta
_{1}^{t},\theta _{2}^{t},\theta _{b},\phi _{b}\}$ is
\begin{equation}
\widetilde{\overline{\mathcal{G}}_{i}^{g}}\equiv \int_{0}^{2\pi
}d\phi \int_{0}^{2\pi }d\Phi _{R}\widetilde{\overline{I}}_{i}^{g}
\end{equation}
and the terms in these expressions only arise from the
helicity-conserving (63,64), and from the helicity-flip (71,72):
\begin{equation}
\widetilde{\overline{\mathcal{G}}_{i}^{g}}|_{0}=4\pi ^{2}[ (\widetilde{\overline{I}}_{+-,+-}^{g}+%
\widetilde{\overline{I}}_{-+,-+}^{g})|_{0}+(\widetilde{\overline{I}}_{++,++}^{g}+\widetilde{\overline{I}}%
_{--,--}^{g})|_{0} ]
\end{equation}
\begin{equation}
\widetilde{\overline{\mathcal{G}}_{i}^{g}}|_{sig}=4\pi ^{2}[ (\widetilde{\overline{I}}%
_{+-,+-}^{g}+\widetilde{\overline{I}}_{-+,-+}^{g})|_{sig}+(\widetilde{\overline{I}}_{++,++}^{g}+%
\widetilde{\overline{I}}%
_{--,--}^{g})|_{sig} ]
\end{equation}

We obtain
\begin{eqnarray}
\widetilde{\overline{\mathcal{G}}^{g}}|_{0} &=&{}\frac{8\pi ^{3}}{3}%
\,\, c(s,\Theta _{B}) \,\, \left\{ \frac{1}{2}\overline{\Gamma
}(0,0)\sin ^{2}\theta _{b}+\overline{\Gamma }(1,1)\sin
^{4}\frac{\theta _{b}}{2}\right\} [\Gamma
(0,0)+\Gamma (-1,-1)] \nonumber \\
&&\left\{ \widetilde{g}_{1}^{g}(s,\Theta _{B})+\overline{{\mathcal{R}}}{%
\mathcal{\,\,}}\widetilde{g}_{2}^{g}(s,\Theta _{B})\cos \theta
_{1}^{t}\cos \theta _{2}^{t}\right\}
\end{eqnarray}
\begin{eqnarray}
\widetilde{\overline{\mathcal{G}}^{g}}|_{sig} &=&-{}\frac{8\sqrt{2}\pi ^{3}}{%
3} \,\, c(s,\Theta _{B}) \,\, \sin \theta _{b}\sin
^{2}\frac{\theta _{b}}{2}\left\{ \overline{\Gamma }_{R}(0,1)\cos
\phi _{b}+\overline{\Gamma }_{I}(0,1)\sin \phi _{b}\right\}
\nonumber \\
&&\lbrack \Gamma (0,0)+\Gamma (-1,-1)]\,\overline{{\mathcal{R}}}{\mathcal{%
\,\,}}\widetilde{g}_{2}^{g}(s,\Theta _{B})\cos \theta _{1}^{t}\sin
\theta _{2}^{t}
\end{eqnarray}
where $\widetilde{g}_{1,2}^{g}(s,\Theta _{B})$ are given in (4,
5).

Finally, if the $\cos \theta _{2}^{t}$ dependence is integrated
out, we obtain the four-angle distribution in \newline $\{\Theta
_{B,}\theta _{1}^{t},\theta
_{b},\phi _{b}\} $%
\begin{eqnarray}
{\overline{\mathcal{G}}_{i}^{g}} \equiv \int_{-1}^{1} d ( \cos
\theta _ {2}^{t} ) \int_{0}^{2\pi }d\phi \int_{0}^{2\pi }d\Phi
_{R}   \widetilde{\overline{I}_{i}^{g}} &=& \int_{-1}^{1} d ( \cos
\theta _{2}^{t} ) \widetilde{\overline{\mathcal{G}}_{i}^{g}}
\end{eqnarray}
which is listed in the introduction section in (21,22).

\section{Derivation of Quark-Production Beam-Referenced \newline Stage-Two
 Spin-Correlation Functions}

 Unlike the treatment of the quark-production contribution in ``I",  we do not integrate out the $ \cos \Theta_B $
 dependence in this paper, as was discussed in the introduction.

 \subsection{ Quark-production density matrix in JW phase convention}

For comparison with the analogous gluon-density-matrix elements
listed in the previous section, including the color factor, the
$t_{1}\overline{t}_{2}$ helicity-conserving quark-production
density-matrix-elements in the $(t\overline{t})_{c.m.}$ system are
\begin{equation}
\rho _{+-,+-}^{q}=\rho _{-+,-+}^{q}=\frac{g^{4}}{9} \,\, (1+\cos
^{2}\Theta _{B})
\end{equation}
\begin{equation}
\rho _{-+,+-}^{q}=\{\rho _{+-,-+}^{q}\}^{\ast }=\frac{g^{4}}{9}
\,\, e^{i2\Phi _{B}} \,\, \sin ^{2}\Theta _{B}
\end{equation}
with an $e^{i2\Phi _{B}}$ factor in (96).  These and the following
density-matrix-elements agree with (56) of ``I".  For
$q_{1}(k)q_{2}(n)\longrightarrow t_{1}(p)\overline{t_{2}}(l)$, $\
s=(k+n)^{2}=(p+l)^{2}=4E^{2},t=(p-k)^{2}=(n-l)^{2}=m^{2}-2E^{2}(1-\beta
\cos \Theta _{B})$, $u=(p-n)^{2}=(k-l)^{2}=m^{2}-2E^{2}(1+\beta
\cos \Theta _{B})$ where $\beta = p/E$. The mixed-helicity
properties quark-production density-matrix-elements are
\begin{eqnarray}
\rho _{-+,++}^{q} &=&\rho _{-+,--}^{q}=-\rho _{--,+-}^{q}=-\rho
_{++,+-}^{q}= \nonumber
\\
\{\rho _{++,-+}^{q}\}^{\ast } &=&\{\rho _{--,-+}^{q}\}^{\ast
}=-\{\rho
_{+-,--}^{q}\}^{\ast }=-\{\rho _{+-,++}^{q}\}^{\ast }= \nonumber \\
&&-\frac{2  mg^{4}}{9s^{1/2}} \,\, e^{i\Phi _{B}} \,\, \sin \Theta
_{B}\cos \Theta _{B}
\end{eqnarray}
with an overall minus sign and $e^{i\Phi _{B}}$ factor. The
helicity-flip quark-production density-matrix-elements are
\begin{equation}
\rho _{++,++}^{q}=\rho _{--,--}^{q}=\rho _{++,--}^{q}=\rho _{--,++}^{q}=%
\frac{4m^{2}g^{4}}{9 s} \,\, \sin ^{2}\Theta _{B}
\end{equation}

These give the quark contributions to the BR-S2SC functions listed
in Section 2 of ``I", when each expression in ``I" is multiplied
by the color factor $\frac{1}{9}$.

\subsection{ Lepton-plus-Jets Channel:  $ \lambda_b = -1/2$,
$\lambda_{\bar{b}} = +1/2$
 \newline Dominance }

\subsubsection{$t_{1}\rightarrow W_{1}^{+}b\rightarrow
(l^{+}\protect\nu )b$}

With the $i=(\lambda _{1}\lambda _{2},\lambda _{1}^{^{\prime
}}\lambda _{2}^{^{\prime }})$ labelling of the present paper, the
quark contributions are proportional to the ones in (98-109) of
``I'' as follows.

For the helicity-conserving contribution,
\begin{equation}
(\widetilde{I}_{+-,+-}^{q}+\widetilde{I}_{-+,-+}^{q})|_{0}=\frac{1}{9}(%
\widetilde{I}_{++}+\widetilde{I}_{--})|_{0}
\end{equation}
\begin{equation}
(\widetilde{I}_{+-,+-}^{q}+\widetilde{I}_{-+,-+}^{q})|_{sig}=\frac{1}{9}(%
\widetilde{I}_{++}+\widetilde{I}_{--})|_{sig}
\end{equation}
\begin{equation}
(\widetilde{I}_{+-,-+}^{q}+\widetilde{I}_{-+,+-}^{q})|_{0}=\frac{1}{9}(%
\widetilde{I}_{+-}+\widetilde{I}_{-+})|_{0}
\end{equation}
\begin{equation}
(\widetilde{I}_{+-,-+}^{q}+\widetilde{I}_{-+,+-}^{q})|_{sig}=\frac{1}{9}(%
\widetilde{I}_{+-}+\widetilde{I}_{-+})|_{sig}
\end{equation}

For the mixed-helicity contribution,
\begin{equation}
(\widetilde{I}_{-+,++}^{q}+\widetilde{I}_{--,+-}^{q}+\widetilde{I}%
_{+-,--}^{q}+\widetilde{I}_{++,-+}^{q})|_{0}=\frac{1}{9}\widetilde{I}^{m(%
\overline{\omega }^{+}+\overline{\eta }^{-})}|_{0}
\end{equation}
\begin{equation}
(\widetilde{I}_{-+,++}^{q}+\widetilde{I}_{--,+-}^{q}+\widetilde{I}%
_{+-,--}^{q}+\widetilde{I}_{++,-+}^{q})|_{sig}=\frac{1}{9}\widetilde{I}^{m(%
\overline{\omega }^{+}+\overline{\eta }^{-})}|_{sig}
\end{equation}
\begin{equation}
(\widetilde{I}_{++,+-}^{q}+\widetilde{I}_{--,-+}^{q}+\widetilde{I}%
_{+-,++}^{q}+\widetilde{I}_{-+,--}^{q})|_{0}=\frac{1}{9}\widetilde{I}^{m(%
\overline{\omega }^{-}+\overline{\eta }^{+})}|_{0}
\end{equation}
\begin{equation}
(\widetilde{I}_{++,+-}^{q}+\widetilde{I}_{--,-+}^{q}+\widetilde{I}%
_{+-,++}^{q}+\widetilde{I}_{-+,--}^{q})|_{sig}=\frac{1}{9}\widetilde{I}^{m(%
\overline{\omega }^{-}+\overline{\eta }^{+})}|_{sig}
\end{equation}

The helicity-flip contributions are
\begin{equation}
(\widetilde{I}_{++,++}^{q}+\widetilde{I}_{--,--}^{q})|_{0}=\frac{1}{9}(%
\widetilde{I}_{++}^{m2}+\widetilde{I}_{--}^{m2})|_{0}
\end{equation}
\begin{equation}
(\widetilde{I}_{++,++}^{q}+\widetilde{I}_{--,--}^{q})|_{sig}=\frac{1}{9}(%
\widetilde{I}_{++}^{m2}+\widetilde{I}_{--}^{m2})|_{sig}
\end{equation}
\begin{equation}
(\widetilde{I}_{++,--}^{q}+\widetilde{I}_{--,++}^{q})|_{0}=\frac{1}{9}(%
\widetilde{I}_{+-}^{m2}+\widetilde{I}_{-+}^{m2})|_{0}
\end{equation}
\begin{equation}
(\widetilde{I}_{++,--}^{q}+\widetilde{I}_{--,++}^{q})|_{sig}=\frac{1}{9}(%
\widetilde{I}_{+-}^{m2}+\widetilde{I}_{-+}^{m2})|_{sig}
\end{equation}

By this $i=(\lambda _{1}\lambda _{2},\lambda _{1}^{^{\prime
}}\lambda _{2}^{^{\prime }})$ labelling, the above general
formulas and remarks (77-83) for the successively fewer-angle
gluon-distributions apply for the quark-distributions by simply
changing the superscript $ g \rightarrow q$.  Thereby, one obtains
the six-angle $\widetilde{\mathcal{H}}_{i}^{q}$,
$(\widetilde{\mathcal{H}}_{i}^{q})^{^{\prime }}$, and the
five-angle  $\widetilde{\mathcal{G}}_{i}^{q}$.

Explicitly, in terms of $\{\Theta _{B},\theta _{1}^{t},\theta
_{2}^{t},\theta _{a},\phi _{a}\}$, the five-angle distribution is
\begin{eqnarray}
\widetilde{\mathcal{G}}^{q}|_{0} &=&\frac{\pi ^{3}g^{4}}{27 s^{2}}
\,\, \left\{
\frac{1}{2}\Gamma (0,0)\sin ^{2}\theta _{a}+\Gamma (-1,-1)\sin ^{4}\frac{%
\theta _{a}}{2}\right\} [\overline{\Gamma }(0,0)+\overline{\Gamma }(1,1)]
\nonumber \\
&&\left\{ \widetilde{g}_{1}^{q}(s,\Theta _{B})+{\mathcal{R\,\,}}\widetilde{g}%
_{2}^{q}(s,\Theta _{B})\cos \theta _{1}^{t}\cos \theta
_{2}^{t}\right\}
\end{eqnarray}
\begin{eqnarray}
\widetilde{\mathcal{G}}^{q}|_{sig} &=&{-}\frac{\sqrt{2}\pi ^{3}g^{4}}{27 s^{2}%
}{}\sin \theta _{a}\sin ^{2}\frac{\theta _{a}}{2} \,\,  \left\{
\Gamma _{R}(0,-1)\cos \phi _{a}-\Gamma _{I}(0,-1)\sin \phi
_{a}\right\}
\nonumber \\
&&[\overline{\Gamma }(0,0)+\overline{\Gamma }(1,1)]\,{\mathcal{R\,\,}}%
\widetilde{g}_{2}^{q}(s,\Theta _{B})\sin \theta _{1}^{t}\cos
\theta _{2}^{t}
\end{eqnarray}
where the two quark-beam-referencing factors
$\widetilde{g}_{1,2}^{q}(s,\Theta _{B})$ are listed in (11, 12).

The simpler four-angle distribution $ {\mathcal{G}}_{i}^{q}  =
\int_{-1}^{1} d ( \cos \theta _{1}^{t} )
\widetilde{\mathcal{G}}_{i}^{q}$ is given in the introduction in
(9, 10).

\subsubsection{$\overline{t}_{2}\rightarrow
W_{2}^{-}\overline{b}\rightarrow (l^{-}\bar{\protect\nu}
)\overline{b}$}

For this CP-conjugate case, the quark contributions are
proportional to (113-124) of ``I''.

For the helicity-conserving contribution,
\begin{equation}
(\widetilde{\overline{I}}_{+-,+-}^{q}+\widetilde{\overline{I}}%
_{-+,-+}^{q})|_{0}=\frac{1}{9}(\widetilde{\overline{I}}_{++}+\widetilde{%
\overline{I}}_{--})|_{0}
\end{equation}
\begin{equation}
(\widetilde{\overline{I}}_{+-,+-}^{q}+\widetilde{\overline{I}}%
_{-+,-+}^{q})|_{sig}=\frac{1}{9}(\widetilde{\overline{I}}_{++}+\widetilde{%
\overline{I}}_{--})|_{sig}
\end{equation}
\begin{equation}
(\widetilde{\overline{I}}_{+-,-+}^{q}+\widetilde{\overline{I}}%
_{-+,+-}^{q})|_{0}=\frac{1}{9}(\widetilde{\overline{I}}_{+-}+\widetilde{%
\overline{I}}_{-+})|_{0}
\end{equation}
\begin{equation}
(\widetilde{\overline{I}}_{+-,-+}^{q}+\widetilde{\overline{I}}%
_{-+,+-}^{q})|_{sig}=\frac{1}{9}(\widetilde{\overline{I}}_{+-}+\widetilde{%
\overline{I}}_{-+})|_{sig}
\end{equation}

For the mixed-helicity contribution,
\begin{equation}
(\widetilde{\overline{I}}_{-+,++}^{q}+\widetilde{\overline{I}}_{--,+-}^{q}+%
\widetilde{\overline{I}}_{+-,--}^{q}+\widetilde{\overline{I}}%
_{++,-+}^{q})|_{0}=\frac{1}{9}\widetilde{\overline{I}}^{m(\overline{\omega }%
^{+}+\overline{\eta }^{-})}|_{0}
\end{equation}
\begin{equation}
(\widetilde{\overline{I}}_{-+,++}^{q}+\widetilde{\overline{I}}_{--,+-}^{q}+%
\widetilde{\overline{I}}_{+-,--}^{q}+\widetilde{\overline{I}}%
_{++,-+}^{q})|_{sig}=\frac{1}{9}\widetilde{\overline{I}}^{m(\overline{\omega
}^{+}+\overline{\eta }^{-})}|_{sig}
\end{equation}
\begin{equation}
(\widetilde{\overline{I}}_{++,+-}^{q}+\widetilde{\overline{I}}_{--,-+}^{q}+%
\widetilde{\overline{I}}_{+-,++}^{q}+\widetilde{\overline{I}}%
_{-+,--}^{q})|_{0}=\frac{1}{9}\widetilde{\overline{I}}^{m(\overline{\omega }%
^{-}+\overline{\eta }^{+})}|_{0}
\end{equation}
\begin{equation}
(\widetilde{\overline{I}}_{++,+-}^{q}+\widetilde{\overline{I}}_{--,-+}^{q}+%
\widetilde{\overline{I}}_{+-,++}^{q}+\widetilde{\overline{I}}%
_{-+,--}^{q})|_{sig}=\frac{1}{9}\widetilde{\overline{I}}^{m(\overline{\omega
}^{-}+\overline{\eta }^{+})}|_{sig}
\end{equation}

For the helicity-flip contribution,
\begin{equation}
(\widetilde{\overline{I}}_{++,++}^{q}+\widetilde{\overline{I}}%
_{--,--}^{q})|_{0}=\frac{1}{9}(\widetilde{\overline{I}}_{++}^{m2}+\widetilde{%
\overline{I}}_{--}^{m2})|_{0}
\end{equation}
\begin{equation}
(\widetilde{\overline{I}}_{++,++}^{q}+\widetilde{\overline{I}}%
_{--,--}^{q})|_{sig}=\frac{1}{9}(\widetilde{\overline{I}}_{++}^{m2}+%
\widetilde{\overline{I}}_{--}^{m2})|_{sig}
\end{equation}
\begin{equation}
(\widetilde{\overline{I}}_{++,--}^{q}+\widetilde{\overline{I}}%
_{--,++}^{q})|_{0}=\frac{1}{9}(\widetilde{\overline{I}}_{+-}^{m2}+\widetilde{%
\overline{I}}_{-+}^{m2})|_{0}
\end{equation}
\begin{equation}
(\widetilde{\overline{I}}_{++,--}^{q}+\widetilde{\overline{I}}%
_{--,++}^{q})|_{sig}=\frac{1}{9}(\widetilde{\overline{I}}_{+-}^{m2}+%
\widetilde{\overline{I}}_{-+}^{m2})|_{sig}
\end{equation}

The above general formulas and remarks in (87-93) for the
successively fewer-angle gluon-distributions apply for the
quark-distributions by simply changing the superscript
$g\rightarrow q$. Thereby, one obtains the six-angle
$\widetilde{\overline{%
\mathcal{H}}_{i}^{q}}$, $(\widetilde{\overline{\mathcal{H}}_{i}^{q}}%
)^{^{\prime }}$, and the five-angle $\widetilde{\overline{\mathcal{G}}%
_{i}^{q}}$.

Explicitly, in terms of $\{\Theta _{B},\theta _{1}^{t},\theta
_{2}^{t},\theta _{b},\phi _{b}\}$, the five-angle distribution is
\begin{eqnarray}
\widetilde{\overline{\mathcal{G}}}^{q}|_{0} &=&\frac{\pi ^{3}g^{4}}{27 s^{2}}%
\,\, \left\{ \frac{1}{2}\overline{\Gamma }(0,0)\sin ^{2}\theta _{b}+\overline{%
\Gamma }(1,1)\sin ^{4}\frac{\theta _{b}}{2}\right\} [\Gamma
(0,0)+\Gamma
(-1,-1)] \nonumber \\
&&\left\{ \widetilde{g}_{1}^{q}(s,\Theta _{B})+\overline{{\mathcal{R}}}{%
\mathcal{\,\,\,}}\widetilde{g}_{2}^{q}(s,\Theta _{B})\cos \theta
_{1}^{t}\cos \theta _{2}^{t}\right\}
\end{eqnarray}
\begin{eqnarray}
\widetilde{\overline{\mathcal{G}}}^{q}|_{sig}
&=&{-}\frac{\sqrt{2}\pi ^{3}g^{4}}{27 s^{2}}{}\sin \theta _{b}\sin
^{2}\frac{\theta _{b}}{2} \,\, \left\{ \overline{\Gamma
}_{R}(0,1)\cos \phi _{b}+\overline{\Gamma }_{I}(0,1)\sin \phi
_{b}\right\} \nonumber \\
&&[\Gamma (0,0)+\Gamma (-1,-1)]\,\,\overline{{\mathcal{R}}}{\mathcal{\,\,\,}}%
\widetilde{g}_{2}^{q}(s,\Theta _{B})\cos \theta _{1}^{t}\sin
\theta _{2}^{t}
\end{eqnarray}

The simpler four-angle distribution
${\overline{\mathcal{G}}_{i}^{q}} = \int_{-1}^{1} d ( \cos \theta
_{2}^{t} ) \widetilde{\overline{\mathcal{G}}_{i}^{q}} $ is given
in the introduction in (24,25).

\section{Summary and Discussion}

From the top-quark beam-referenced spin-correlation function ${
\mathcal{G}}^{(g,q)} {|} _{0} + {\mathcal{G}}^{(g,q)} {|}_{sig} $
with both gluon (2,3) and quark (9,10) production contributions,
the tests for $t_1 \rightarrow {W}^+ b $ decay are:

(i) By measurement of ${\Gamma }_{R}(0,-1)$, the relative sign of
the two dominant $\lambda_b=-1/2$ helicity-amplitudes can be
determined if their relative phase is $0^0$ or $180^0$. Versus the
partial-decay-width $\Gamma ( t \rightarrow W^+ b ) $, W-boson
longitudinal-transverse interference is a large effect because in
the standard model $\eta_L \equiv \frac{{\Gamma
}_{R}(0,-1)}{\Gamma} = \pm 0.46 $ without/with a large $ t_R
\rightarrow b_L $ chiral weak-transition-moment. [ $\Gamma \equiv
$ partial width for $t \rightarrow W^+ b$ ]

(ii) By measurement of both ${\Gamma }_{R}(0,-1)$ and ${\Gamma
}_{I}(0,-1)$ via the $\phi_a$ dependence, limits can be set on a
possible non-trivial phase $\beta _{L}  \equiv  \varphi
_{-1,-\frac{1}{2}}-\varphi _{0,-\frac{1}{2}}$, see (14-16).
Non-trivial relative phases can occur in top-quark decays if
$\widetilde T_{FS}$ invariance is violated [7,1]. Such a violation
will occur if either (a) there is a fundamental violation of
canonical time-reversal invariance, and/or (b) there are
absorptive final-state interactions.

Explicit expressions for the $A(\lambda _{W^{+}},\lambda _{b})$
helicity amplitudes in terms of the most general Lorentz coupling
\begin{eqnarray}
W_\mu ^{*} J_{\bar b t}^\mu = W_\mu ^{*}\bar u_{b}\left( p\right)
\Gamma ^\mu u_t \left( k\right)
\nonumber
\end{eqnarray}
where $k_t =q_W +p_b $, are given in the NKLM paper in [7]. The
canonical decomposition of $\Gamma ^\mu  = {\Gamma ^\mu}_V +
{\Gamma ^\mu}_A $ is
\begin{eqnarray}
\Gamma _V^\mu =g_V\gamma ^\mu + \frac{f_M}{2\Lambda_M }\iota
\sigma ^{\mu \nu }(k-p)_\nu + \frac{g_{S^{-}}}{2\Lambda_{S^{-}}
}(k-p)^\mu \nonumber \\ +\frac{g_S}{2\Lambda_S }(k+p)^\mu
+%
\frac{g_{T^{+}}}{2\Lambda_{T^{+}} }\iota \sigma ^{\mu \nu
}(k+p)_\nu \nonumber
\end{eqnarray}
\begin{eqnarray}
\Gamma _A^\mu =g_A\gamma ^\mu \gamma _5+ \frac{f_E}{2\Lambda_E
}\iota \sigma ^{\mu \nu }(k-p)_\nu \gamma _5 +
\frac{g_{P^{-}}}{2\Lambda_{P^{-}} }(k-p)^\mu \gamma _5  \nonumber \\
+\frac{g_P}{2\Lambda_P }%
(k+p)^\mu \gamma _5  +\frac{g_{T_5^{+}}}{2\Lambda_{T_5^{+}} }\iota
\sigma ^{\mu \nu }(k+p)_\nu \gamma _5 \nonumber
\end{eqnarray}
where the parameters $\Lambda_i =$ ``effective-mass scale of new
physics associated with the $i$th type additional Lorentz
structure." For a general treatment of additional Lorentz
structures to pure $(V-A)$, the $g_i$ or $\Lambda_i$ must be
considered as complex phenomenological parameters. Details, such
as Lorentz-structure-equivalence theorems $S \sim V + f_M$, $P
\sim -A + f_E, \ldots$ ; the matrix elements of the divergences of
these couplings; and the definitions of the chiral couplings
$g_{L,R} = g_V \mp g_A, \ldots$ are in the NKLM, NC, NA papers in
[7].

For $\bar{t}_2 \rightarrow {W}^- \bar{b} $ decay, from the gluon
(21,22) and quark (24,25) production contributions to the
CP-conjugate BR-S2SC function for $ \overline{{ \mathcal{G}}^{g,q}
{|}} _{0} + \overline{{\mathcal{G}}^{g,q} {|}}_{sig} $, there are
two analogous tests for $\bar{t}_2 \rightarrow {W_2}^- \bar{b} $
decay.  By measurement of $\overline{\Gamma }_{R}(0,1)$, the
relative sign of the two dominant helicity amplitudes $B(0,1/2)$
and $B(1,1/2)$ for $\bar{t}_2 \rightarrow {W}^- \bar{b} $ can be
determined if their relative phase is $0^0$ or $180^0$ ( as in the
case of large $ \bar{t}_L \rightarrow \bar{b}_R $ chiral
weak-transition-moment).  By measurement of both $\overline{\Gamma
}_{R}(0,1)$ and $\overline{\Gamma }_{I}(0,1)$ via the $\phi_b$
dependence, limits can be set on a non-trivial phase
$\overline{\beta }_{R}  \equiv \overline{\varphi }_{1,\frac{1}{2}}-\overline{%
\varphi }_{0,\frac{1}{2}}$, see (18-20).

In all these BR-S2SC functions for top-quark decay tests, the
polarized-partial-widths and W-boson-LT-interference-widths
(15,16) appear multiplied by the   $\theta_a$, $\phi_a$ angular
factors, which are expected in the helicity-formalism for the
decay chain $t_1 \rightarrow W^{+} b \rightarrow ( l^+ \nu ) b $.
The spherical angles $ \theta_a$, $\phi_a $ specify the $ l^+ $
momentum in the ${W_1}^+$ rest frame when there is first a boost
from the $ (t \bar{t} )_{c.m.}$ frame to the $t_1$ rest frame, and
then a second boost from the $t_1$ rest frame to the ${W_1}^+$
rest frame. So, $\frac{1}{2}\Gamma (0,0)\sin ^{2}\theta _{a}$ and
$\Gamma (-1,-1)\sin ^{4}\frac{\theta _{a}}{2}$  appear in the
background terms ${\mathcal{G}}^{(g,q)} {|} _{0}$, see (2, 9).
Similarly, $ \Gamma _{R}(0,-1) \sin \theta _{a} \sin
^{2}\frac{\theta _{a}}{2} \cos \phi _{a}$  and $ \Gamma _{I}(0,-1)
\sin \theta _{a} \sin ^{2}\frac{\theta _{a}}{2} \sin \phi _{a}$
appear in the signal terms ${\mathcal{G}}^{(g,q)} {|} _{sig}$, see
(3, 10). The situation is analogous for the $ \theta_b$, $\phi_b $
variables in the CP-conjugate BR-S2SC functions.

The above summary is for the leading-order in QCD and the
leading-order in electroweak interactions considered in this
paper, assuming that the $\lambda_b=-1/2$ and $\lambda_{\bar{b}}=
 1/2$ helicity-amplitudes dominate in $t \rightarrow W^+ b$ decay.
An important consequence of such dominance is that the W-boson
longitudinal-transverse-interference effects and BR-S2SC
signatures treated in this paper are both large versus the
non-dominant contributions and versus higher-order QCD and/or
higher-order electroweak contributions. However, for later
measurement, there are the two $\lambda_b = 1/2$ non-dominant
amplitudes in $t_1 \rightarrow {W}^+ b $ decay with their two
moduli and two additional relative phases.  For a clear and simple
``visual display" of these measurable phases, see
``$\alpha,\beta,\gamma$"-relative-phases in Fig. 1 in the NC paper
in [7]. In this context, next-to-leading-order QCD,
next-to-leading-order electroweak, and also W-boson and t-quark
finite-width corrections require further theoretical investigation
[8].  Other polarimetry techniques such as $\Lambda_b$ polarimetry
[9], and excellent understandings of detector-systematics and of
reaction-backgrounds will be required for a complete measurement
of the four $t \rightarrow W^+ b$ decay amplitudes and of the four
decay-amplitudes for the CP-conjugate mode.

One learns a number of things from this derivation of the
production density matrices and these associated BR-S2SC
functions.  First, the non-trivial overall minus signs and
$e^{{\pm} i \Phi_B}$, $e^{{\pm} i 2  \Phi_B}$ factors appear in
the same patterns in the gluon and quark production density
matrices. This has the important and empirically useful
consequence that most relative phase effects in these BR-S2SC
functions do not depend upon whether the final $t_1 \bar{t}_2$
system has been produced by gluon or by quark production.
Therefore, as discussed in Sec. 1.1, there is a common final-state
interference structure of the four-angle BR-S2SC functions, and
likewise for the other additional-angle generalizations which are
listed in Sections 2 and 3.  This is a common interference
structure for W-boson longitudinal-transverse-interference.  When
the $\phi$-dependence is included, there is also a common
interference-structure for $t_1$-quark left-right helicity
interference and for $\bar{t}_2$-antiquark left-right helicity
interference. This left-right spin-$1/2$ interference is in the
gluon contribution, see (61-2), and in the quark contribution, see
(109-110) in this paper and in (108-9) in ``I".

Second, for the $t\rightarrow W^{+}b$ decay mode, one learns that
for a spin-correlation measurement using the ``lepton + jets decay
channel" to determine the relative sign of or to obtain a
constraint on a possible non-trivial phase between the two
dominant $\lambda_b = -1/2 $ helicity amplitudes requires use of
the $ ( t \bar{t} )_{c.m.} $ energy of the hadronically decaying
W-boson, or the kinematically equivalent cosine of the polar-angle
of $W^{\mp}$ emission in the anti-top (top) decay frame. Both this
$ \cos \theta _{2}^{t} $ factor and the $ {\mathcal{R}} =
({\mathtt{prob}} \, W_L) - ({\mathtt{prob}} \, W_T) $ suppression
factor appear in all the four-angle and five-angle signal terms,
but not in any of the background terms. For this application of
W-boson longitudinal-transverse-interference, it is fortunate that
in the standard model the probabilities for the presence of
longitudinal/transverse W-bosons are both large.  Because of this
${\mathcal{R}}$ factor, it is also indeed fortunate that these
probabilities are unequal, $P(W_{L}) = {{\Gamma }(0,0)} / {\Gamma}
= 0.70$ and $P(W_{T}) = {{\Gamma }(-1,-1)} / {\Gamma} = 0.30$
irrespective of whether there exists a large $t_R \rightarrow b_L$
chiral weak-transition moment.

\begin{center}
{\bf Acknowledgments}
\end{center}

This work was partially supported by U.S. Dept. of Energy Contract
No. DE-FG 02-86ER40291.

\begin{appendix}

\section{Appendix:  Spinors and Their Outer-Products in JW Phase Convention}

In the JW phase-convention, for application to
$t_{1}(p)\overline{t}_2 (l)$
pair production with respective helicities $\lambda _{1}$ and $\lambda _{2}$%
,\ the  first particle $p^{\mu}$ spinor outer-products
$u(p,\lambda _{1} )\overline{u}(p,{\lambda _{1}}^{\prime} ),
\ldots $ [10-12] are
\begin{eqnarray}
u(p,\pm )\overline{u}(p,\pm ) &=&\frac{1}{2}(\not{p}+m)(1\pm \gamma _{5}\not%
{S})=\frac{1}{2}(1\pm \gamma _{5}\not{S})(\not{p}+m) \nonumber \\
u(p,+)\overline{u}(p,-) &=&\frac{1}{2}e^{\iota \phi }(\not{p}+m)\gamma _{5}%
\not{C}=\frac{1}{2}e^{\iota \phi }\gamma _{5}\not{C}(\not{p}+m) \nonumber \\
u(p,-)\overline{u}(p,+) &=&\frac{1}{2}e^{-\iota \phi }(\not{p}+m)\gamma _{5}%
\not{C}^{\ast }=\frac{1}{2}e^{-\iota \phi }\gamma _{5}\not{C}^{\ast }(\not%
{p}+m)      \hskip 30mm {(A.1)} \nonumber
\end{eqnarray}
with
\begin{eqnarray}
p^{\mu } &=&(E;p\sin \theta \cos \phi ,p\sin \theta \sin \phi
,p\cos \theta
),\,p^{2}=m^{2} \nonumber \\
S^{\mu } &=&(\frac{p}{m};\frac{E}{m}\hat{p}),\,S^{2}=-1  \nonumber \\
C^{\mu } &=&(0;\cos \theta \cos \phi -\iota \sin \phi ,\cos \theta
\sin \phi +\iota \cos \phi ,-\sin \theta ),\,C\cdot C^{\ast }=-2
\hskip 15mm {(A.2)} \nonumber
\end{eqnarray}
with $\iota=\sqrt{-1}$ and where the asterisk denotes complex
conjugation. The signs of $\lambda _{1}$ and $\lambda _{2}$ are
used in labelling the spinors.  Note that $S\cdot p=0$ and
$C\cdot p=C^{\ast }\cdot p=C\cdot S=C^{\ast }\cdot S=0$.  For
completeness, if the first particle were an anti-particle, then
\begin{eqnarray}
v(p,\pm )\overline{v}(p,\pm ) &=&\frac{1}{2}(\not{p}-m)(1\pm \gamma _{5}\not%
{S})=\frac{1}{2}(1\pm \gamma _{5}\not{S})(\not{p}-m) \nonumber \\
v(p,+)\overline{v}(p,-) &=&\frac{1}{2}e^{\iota \phi }(\not{p}-m)\gamma _{5}%
\not{C}^{\ast }=\frac{1}{2}e^{\iota \phi }\gamma _{5}\not{C}^{\ast }(\not%
{p}-m) \nonumber \\
v(p,-)\overline{v}(p,+) &=&\frac{1}{2}e^{-\iota \phi }(\not{p}-m)\gamma _{5}%
\not{C}=\frac{1}{2}e^{-\iota \phi }\gamma _{5}\not{C}(\not{p}-m)
\hskip 30mm {(A.3)} \nonumber
\end{eqnarray}

The second anti-particle $l^{\mu}$ spinor outer-products
$v(l,{\lambda _{2}}^{\prime} )\overline{v}(l,\lambda _{2} ),
\ldots $ are
\begin{eqnarray}
v(l,\pm )\overline{v}(l,\pm ) &=&\frac{1}{2}(\not{l}-m)(1\pm \gamma _{5}\not%
{R})=\frac{1}{2}(1\pm \gamma _{5}\not{R})(\not{l}-m) \nonumber \\
v(l,+)\overline{v}(l,-) &=&\frac{1}{2}e^{-\iota \phi }(\not{l}-m)\gamma _{5}%
\not{B}=\frac{1}{2}e^{-\iota \phi }\gamma _{5}\not{B}(\not{l}-m) \nonumber \\
v(l,-)\overline{v}(l,+) &=&\frac{1}{2}e^{\iota \phi }(\not{l}-m)\gamma _{5}%
\not{B}^{\ast }=\frac{1}{2}e^{\iota \phi }\gamma _{5}\not{B}^{\ast }(\not%
{l}-m)  \hskip 30mm {(A.4)} \nonumber
\end{eqnarray}
with
\begin{eqnarray}
l^{\mu } &=&(E;-l\sin \theta \cos \phi ,-l\sin \theta \sin \phi
,-l\cos
\theta ),\,l^{2}=m^{2} \nonumber \\
R^{\mu } &=&(\frac{l}{m};-\frac{E}{m}\hat{l}),\,R^{2}=-1 \nonumber \\
B^{\mu } &=&(0;\cos \theta \cos \phi -\iota \sin \phi ,\cos \theta
\sin \phi +\iota \cos \phi ,-\sin \theta ),\,B\cdot B^{\ast }=-2
\hskip 15mm {(A.5)} \nonumber
\end{eqnarray}
so $R\cdot l=0$ and $B\cdot l=B^{\ast }\cdot l=B\cdot R=B^{\ast
}\cdot R=0$.
\ In the  $(t_{1}\overline{t}_{2})_{c.m.}$ system, where the  $\overline{%
t}_{2}(l)$ is back-to-back with the  $t_{1}(p)$, $C\cdot l=C^{\ast
}\cdot l=C\cdot R=C^{\ast }\cdot R=0$. \ In this system, $B^{\mu
}$ and $C^{\mu }$ are identical but in calculating, it is
sometimes useful to keep them distinct.  For completeness, if
$l^{\mu }$ were a particle then
\begin{eqnarray}
u(l,\pm )\overline{u}(l,\pm ) &=&\frac{1}{2}(\not{l}+m)(1\pm \gamma _{5}\not%
{R})=\frac{1}{2}(1\pm \gamma _{5}\not{R})(\not{l}+m) \nonumber \\
u(l,+)\overline{u}(l,-) &=&\frac{1}{2}e^{-\iota \phi }(\not{l}+m)\gamma _{5}%
\not{B}=\frac{1}{2}e^{-\iota \phi }\gamma _{5}\not{B}(\not{l}+m) \nonumber \\
u(l,-)\overline{u}(l,+) &=&\frac{1}{2}e^{\iota \phi }(\not{l}+m)\gamma _{5}%
\not{B}^{\ast }=\frac{1}{2}e^{\iota \phi }\gamma _{5}\not{B}^{\ast }(\not%
{l}+m) \hskip 25mm {(A.6)} \nonumber
\end{eqnarray}

These outer-products follow from the following spinors
$u(p,\lambda )|_{\phi }$, $v(p,\lambda )|_{\phi }$, $\ldots$
constructed following the procedure of [12].  The sub-label $\phi$
on these spinors denotes the non-zero value of the third-Euler
angle in the Wigner $D$-function, see discussion below. This
sub-label is normally suppressed because it is apparent from the
context. For the first particle $p^{\mu }$,
\begin{eqnarray*}
u(p,+)=\frac{1}{\sqrt{E+m}}(\not{p}+m)\left[
\begin{array}{c}
\cos \frac{\theta }{2} \\
e^{i\phi }\sin \frac{\theta }{2} \\
0 \\
0
\end{array}
\right] =\sqrt{E+m}\left[
\begin{array}{c}
\cos \frac{\theta }{2} \\
e^{i\phi }\sin \frac{\theta }{2} \\
\frac{p}{E+m}\left[
\begin{array}{c}
\cos \frac{\theta }{2} \\
e^{i\phi }\sin \frac{\theta }{2}
\end{array}
\right]
\end{array}
\right] \nonumber
\end{eqnarray*}
\begin{eqnarray*}
u(p,-)=\frac{1}{\sqrt{E+m}}(\not{p}+m)\left[
\begin{array}{c}
-e^{-i\phi }\sin \frac{\theta }{2} \\
\cos \frac{\theta }{2} \\
0 \\
0
\end{array}
\right] =\sqrt{E+m}\left[
\begin{array}{c}
-e^{-i\phi }\sin \frac{\theta }{2} \\
\cos \frac{\theta }{2} \\
\frac{p}{E+m}\left[
\begin{array}{c}
e^{-i\phi }\sin \frac{\theta }{2} \\
-\cos \frac{\theta }{2}
\end{array}
\right]
\end{array}
\right]
\end{eqnarray*}
\begin{eqnarray*}
v(p,+)=-\frac{1}{\sqrt{E+m}}(\not{p}-m)\left[
\begin{array}{c}
0 \\
0 \\
\sin \frac{\theta }{2} \\
-e^{i\phi }\cos \frac{\theta }{2}
\end{array}
\right] =-\sqrt{E+m}\left[
\begin{array}{c}
\frac{p}{E+m}\left[
\begin{array}{c}
-\sin \frac{\theta }{2} \\
e^{i\phi }\cos \frac{\theta }{2}
\end{array}
\right] \\
\sin \frac{\theta }{2} \\
-e^{i\phi }\cos \frac{\theta }{2}
\end{array}
\right]
\end{eqnarray*}
\begin{eqnarray}
v(p,-)=-\frac{1}{\sqrt{E+m}}(\not{p}-m)\left[
\begin{array}{c}
0 \\
0 \\
e^{-i\phi }\cos \frac{\theta }{2} \\
\sin \frac{\theta }{2}
\end{array}
\right] =-\sqrt{E+m}\left[
\begin{array}{c}
\frac{p}{E+m}\left[
\begin{array}{c}
e^{-i\phi }\cos \frac{\theta }{2} \\
\sin \frac{\theta }{2}
\end{array}
\right] \\
e^{-i\phi }\cos \frac{\theta }{2} \\
\sin \frac{\theta }{2}
\end{array}
\right] \hskip 5mm {(A.7)} \nonumber
\end{eqnarray}
For the second particle specified by $l^{\mu }$ of (A.5), the
spinors are
\begin{eqnarray*}
u(l,-)=\frac{1}{\sqrt{E+m}}(\not{l}+m)\left[
\begin{array}{c}
\cos \frac{\theta }{2} \\
e^{i\phi }\sin \frac{\theta }{2} \\
0 \\
0
\end{array}
\right] =\sqrt{E+m}\left[
\begin{array}{c}
\cos \frac{\theta }{2} \\
e^{i\phi }\sin \frac{\theta }{2} \\
-\frac{l}{E+m}\left[
\begin{array}{c}
\cos \frac{\theta }{2} \\
e^{i\phi }\sin \frac{\theta }{2}
\end{array}
\right]
\end{array}
\right]
\end{eqnarray*}
\begin{eqnarray*}
u(l,+)=\frac{1}{\sqrt{E+m}}(\not{l}+m)\left[
\begin{array}{c}
-e^{-i\phi }\sin \frac{\theta }{2} \\
\cos \frac{\theta }{2} \\
0 \\
0
\end{array}
\right] =\sqrt{E+m}\left[
\begin{array}{c}
-e^{-i\phi }\sin \frac{\theta }{2} \\
\cos \frac{\theta }{2} \\
\frac{l}{E+m}\left[
\begin{array}{c}
-e^{-i\phi }\sin \frac{\theta }{2} \\
\cos \frac{\theta }{2}
\end{array}
\right]
\end{array}
\right]
\end{eqnarray*}
\begin{eqnarray*}
v(l,-)=-\frac{1}{\sqrt{E+m}}(\not{l}-m)\left[
\begin{array}{c}
0 \\
0 \\
\sin \frac{\theta }{2} \\
-e^{i\phi }\cos \frac{\theta }{2}
\end{array}
\right] =-\sqrt{E+m}\left[
\begin{array}{c}
\frac{l}{E+m}\left[
\begin{array}{c}
\sin \frac{\theta }{2} \\
-e^{i\phi }\cos \frac{\theta }{2}
\end{array}
\right] \\
\sin \frac{\theta }{2} \\
-e^{i\phi }\cos \frac{\theta }{2}
\end{array}
\right]
\end{eqnarray*}
\begin{eqnarray}
v(l,+)=-\frac{1}{\sqrt{E+m}}(\not{l}-m)\left[
\begin{array}{c}
0 \\
0 \\
e^{-i\phi }\cos \frac{\theta }{2} \\
\sin \frac{\theta }{2}
\end{array}
\right] =-\sqrt{E+m}\left[
\begin{array}{c}
-\frac{l}{E+m}\left[
\begin{array}{c}
e^{-i\phi }\cos \frac{\theta }{2} \\
\sin \frac{\theta }{2}
\end{array}
\right] \\
e^{-i\phi }\cos \frac{\theta }{2} \\
\sin \frac{\theta }{2}
\end{array}
\right]  \hskip 5mm {(A.8)} \nonumber
\end{eqnarray}

A useful alternative spinor-construction is to set the third-Euler
angle to zero.  When needed for clarity, we denote this alternate
spinor-construction by the sub-label ``0".  The formulas for these
$u(p,\lambda )|_{0}$ and $v(p,\lambda )|_{0}$ spinors are
proportional to those listed above. Explicitly, for $p^{\mu}$,
\begin{eqnarray}
u(p,\pm )|_{0}=e^{\mp i\phi /2}\,u(p,\pm )|_{\phi },\,\,\,v(p,\pm
)|_{0}=e^{\mp i\phi /2}\,v(p,\pm )|_{\phi } \hskip 25mm (A.9)
\nonumber
\end{eqnarray}
and for $l^{\mu}$,
\begin{eqnarray}
u(l,\pm )|_{0}=e^{\pm i\phi /2}\,u(l,\pm )|_{\phi },\,\,\,v(l,\pm
)|_{0}=e^{\pm i\phi /2}\,v(l,\pm )|_{\phi } \hskip 28mm {(A.10)}
\nonumber
\end{eqnarray}
Consequently, the $e^{\pm i\phi }$ factors are absent in their
associated outer-products, so in place of the above
outer-products:
$\, \, u(p,+)|_{0}\,\overline{u}(p,-)|_{0}=%
\frac{1}{2}(\not{p}+m)\gamma _{5}\not{C},$ $\ldots$. Similarly, in
place of the spin-one particle $\epsilon ^{\mu }(p,\lambda
)|_{\phi }$ polarization vectors in [11],
\begin{eqnarray}
\epsilon ^{\mu }(p,\lambda )|_{0}=e^{-i\,\lambda \,\phi
}\,\,\epsilon ^{\mu }(p,\lambda )|_{\phi }\,;\,\,\lambda =\pm 1,0
\hskip 35mm {(A.11)} \nonumber
\end{eqnarray}

The $\mathbf{P}$ parity and $\mathbf{T}$ time-reversal discrete
symmetry properties for the above spinors do not depend on the
choice of the third Euler angle. With $\mathbf{P=\gamma }%
^{0}$ and $\lambda =\pm $%
\begin{eqnarray*}
{ \mathbf{P} } u(p,\lambda ) &=&u(l,-\lambda )  \\
{ \mathbf{P} }  v(p,\lambda ) &=&- v(l,-\lambda )
\end{eqnarray*}
where $l=|\overrightarrow{l}|=p=|\overrightarrow{p}|$ . In these
and the following relations, the $p$ denotes the $u(p,\lambda
)|_{\phi }$ spinor which is listed in (A.7) and the $l$ denotes
the $u(l,\lambda )|_{\phi }$ spinor which is listed in (A.8), etc.
. The gamma matrices are in the Dirac representation, which is
sometimes named the ``standard" or ``Dirac-Pauli" [12]
representation.
With $\mathbf{T}=$ $i$ $\mathbf{\gamma }^{2}\mathbf{\gamma }^{5}\mathbf{%
\gamma }^{0}$ $= \mathbf{\gamma }^{1}\mathbf{\gamma }^{3}$
\begin{eqnarray*}
{ \mathbf{T} } u^{\ast }(p,\pm ) &=&\mp u(l,\pm )  \\
{ \mathbf{T} } v^{\ast }(p,\pm ) &=&\mp v(l,\pm )
\end{eqnarray*}
where $l=|\overrightarrow{l}|=p=|\overrightarrow{p}|$, and the
asterisk denotes complex conjugation.

The $\mathbf{C}$ charge-conjugation discrete symmetry properties
do depend
on the choice of the third Euler angle.  With $\mathbf{C=}$ $i$ $\mathbf{%
\gamma }^{2}\mathbf{\gamma }^{0}$ , for the first particle $p^{\mu
}$ spinors (A.7)
\begin{eqnarray*}
v(p,\pm )|_{\phi } &=&e^{\pm i\phi } i  {\gamma }%
^{2}\,u^{\ast }(p,\pm )|_{\phi }=e^{\pm i\phi } { \mathbf{C} } %
\overline{u}^{T}(p,\pm )|_{\phi }  \\
u(p,\pm )|_{\phi } &=&e^{\pm i\phi } i {\gamma }%
^{2}\,v^{\ast }(p,\pm )|_{\phi }
\end{eqnarray*}
and for the second particle $l^{\mu }$ spinors\bigskip\ (A.8)
\begin{eqnarray*}
v(l,\pm )|_{\phi } &=&e^{\mp i\phi } i {\gamma }%
^{2}\,u^{\ast }(l,\pm )|_{\phi }=e^{\mp i\phi } { \mathbf{C} } %
\overline{u}^{T}(l,\pm )|_{\phi }
 \\
u(l,\pm )|_{\phi } &=&e^{\mp i\phi } i {\gamma }%
^{2}\,v^{\ast }(l,\pm )|_{\phi }
\end{eqnarray*}
The $\mathbf{C}$ relations for the $u(p,\pm )|_{0}\ $and $v(p,\pm
)|_{0}$ spinors are obtained by omitting the $e^{\pm i\phi }$
factors in these equations.

In the gluon (quark) production density matrices used in the text,
we follow the $u(p,\lambda )|_{\phi}$, $v(p,\lambda )|_{\phi}$
choice which corresponds to the choice made in ``I'' to use a
non-zero third-Euler angle in describing $t_1 \bar{t}_2$ pair
production in the helicity-formalism. Following the Jacob-Wick
papers [5], this non-zero choice of third-argument in the various
Wigner $D$-functions is more often chosen in the literature on
applications of the helicity-formalism. An important exception
occurs in treating sequential decay chains. For instance, in ``I"
for describing the sequential-decay matrix $R_{\lambda _1\lambda
_1^{^{\prime }}}$ for $t\rightarrow W^{+}b\rightarrow (l^{+}\nu
)b$ in ``I'', we do set the third-Euler angle equal to zero. For
specifying the second-stage axes orientation for $W^{+}
\rightarrow l^{+}\nu $ versus the first-stage axes from
$t\rightarrow W^{+}b$ where the $W^+$-momentum is in the
$D$-functions, it is conceptually simpler not to make an
unnecessary additional rotation about the $W^{+}$-axis in the
first-stage $D$-functions, see Sec. II of [13]. For $t\rightarrow
W^{+}b$ decay, unlike in completely orienting a rigid body in
classical mechanics, it is only necessary to correctly orient the
$\theta _{1}^{t},\phi _{1}^{t}$ direction of the $W^{+}$-momentum.
This only involves the first two Euler angles. Unlike for
specifying the $W^{+}$-momentum, in the case of orienting a rigid
body, a third Euler angle rotation is required for one must also
specify the amount of angular rotation of the
non-cylindrically-symmetric rigid body about its $\theta
_{1}^{t},\phi _{1}^{t}$ axis.

Obviously, some care is needed to insure that the choice of
spinors corresponds to the choice of third argument used in the
$D$-functions in the helicity-formalism.  This occasionally needs
emphasis because it is common to use the $u(p,\lambda )|_{0}$ and
$ v(p,\lambda )|_{0}$ type spinors, and also common to use
non-zero third-argument $D$-functions in the helicity-formalism.
If used simultaneously to describe the same elementary particle
reaction, this would be an inconsistent treatment of
relative-phase effects.

For both helicity choices, these spinors were checked by using
them to calculate decay matrix elements and comparing their
overall signs and $\theta$, $\phi$ dependence with those of the
helicity-formalism: Using the spinors for the final
neutrinos/anti-neutrinos, the first particle $p^{\mu}$ spinors
were checked for $\tau^- \rightarrow \nu \pi^-$, $\tau^+
\rightarrow \bar{\nu} \pi^+$ and the second particle $l^{\mu}$
spinors were checked for $\tau^- \rightarrow \pi^- \nu $, $\tau^+
\rightarrow \pi^+ \bar{\nu}$. They were checked for $Z^o
\rightarrow \tau^- \tau^+$ and $Z^o \rightarrow \tau^+ \tau^- $.
The quark-production density-matrix elements for $q_1 \bar{q}_2
\rightarrow t_1 \bar{t}_2$ were calculated as in Appendix B and
agree with eq.(56) of ``I".

\section{Appendix: Derivation of Gluon-Production \newline
Density-Matrix-Elements in JW Phase Convention}

For the gluon-production sequence $ g_1 g_2 \rightarrow t_1
\overline{t}_2 $, the density-matrix-elements are
\begin{eqnarray}
\rho _{\lambda _{1}\lambda _{2},\lambda _{1}^{^{\prime }}\lambda
_{2}^{^{\prime }}}^{g}(\Theta _{B},\Phi _{B})=\frac{1}{4}\sum_{s_{1},s_{2}}%
{\mathcal{M}}(s_{1}s_{2},\lambda _{1}\lambda
_{2})\,{\mathcal{M}}^{* }(s_{1}s_{2},\lambda _{1}^{^{\prime
}}\lambda _{2}^{^{\prime }}) \hskip 25mm {(B.1)} \nonumber
\end{eqnarray}
for the QCD Lorentz-invariant amplitude $\mathcal{M}$ to the
leading-order in $\alpha _{s}$. The initial gluon spin-averaging
 is over the massless $g_{1}$ and $g_{2}$ helicities
  $s_{1}$, $s_{2}$. The $t_{1}$ and
$\overline{t}_{2}$ helicites in the amplitude are $\lambda _{1}$,
$ \lambda _{2}$.  In the complex-conjugate amplitude, $\lambda
_{1}^{^{\prime }}$, $\lambda _{2}^{^{\prime }}$ are their
helicities. For simplicity, the initial color-averaging, the final
color-summation, and the associated color-indices are not
displayed explicitly in (B.1).   With the spinor outer-products of
Appendix A, from (B.1) by normal tracing techniques, the
expressions given in the text follow for $\rho _{\lambda
_{1}\lambda _{2},\lambda _{1}^{^{\prime }}\lambda _{2}^{^{\prime
}}}^{g}$ and analogously for $\rho _{\lambda _{1}\lambda
_{2},\lambda _{1}^{^{\prime }}\lambda _{2}^{^{\prime }}}^{q}$,
compare Appendix A of [14]. Similarly, for the $\Theta _{t},\Phi
_{t}$ beam coordinate system of  Fig. 3, the alternative
$\widetilde{\rho }_{\lambda _{1}\lambda _{2},\lambda
_{1}^{^{\prime }}\lambda _{2}^{^{\prime }}}^{g}$ gluon production
density-matrix-elements of Appendix C follow. For either beam
coordinate system, there are 5 distinct gluon-production
density-matrix-elements and 5 distinct quark-production
density-matrix-elements. The other density-matrix-elements follow
either by hermiticity or by the $P$ and $C$ discrete symmetries
which are present in the outer-products and in the QCD
Feynman-rules.

\section{Appendix:  $\Theta _{t}$ , $\Phi_{t}$ Production
Density-Matrix-Elements in JW Phase Convention}

As shown in Fig. 3, the alternative $\Theta_t,\Phi_t$
beam-referenced production density-matrices $ {
\widetilde{\rho}_{\lambda _{1}\lambda _{2};\lambda _{1}^{^{\prime
}}\lambda _{2}^{^{\prime }}}} $ are for a beam coordinate system $
( {x_b}, {y_b}, {z_b} ) $ in the $ (t \bar{t} )_{c.m.}$ frame in
which the $g_1$ gluon-momentum or $q_1$ quark-momentum ``beam"
direction defines the positive $z_b$ axis. The final $t_1$
momentum is specified by the spherical angles $\Theta_t, \Phi_t$.
In this appendix, we denote these alternative
density-matrix-elements by a ``tilde".

We do not use these alternative density-matrix-elements in the
derivation of the BR-S2SC's because the azimuthal-angle $\Phi_t$
is specified versus the initial-beam direction, whereas in $
{\rho}_{\lambda _{1}\lambda _{2};\lambda _{1}^{^{\prime }}\lambda
_{2}^{^{\prime }}} $ derived in Appendix B and used in the text,
the azimuthal-angle $\Phi_B$ is defined versus the final $t_1$
momentum direction in the $(t\bar{t})_{c.m.}$.  In using the
helicity-formalism for BR-S2SC functions for two-body pair
production such as $t_1 \bar{t_2}$, the final $t_1$ momentum
direction is the natural/most-convenient axis [14] for analyzing
the two sequential-decay-chain-processes and, therefore, for
simultaneously incorporating the beam-referencing.

In the JW phase convention, the  gluon-production
density-matrix-elements are
\begin{eqnarray}
\widetilde{\rho }_{+-,+-}^{g}=\widetilde{\rho }_{-+,-+}^{g}=\widetilde{c}%
(s,\Theta _{t})  \,\, \frac{4p^{2}}{s} \,\, \sin ^{2}\Theta
_{t}(1+\cos ^{2}\Theta _{t}) \hskip 25mm {(C.1)} \nonumber
\end{eqnarray}
\begin{eqnarray}
\widetilde{\rho }_{-+,+-}^{g}=\{\widetilde{\rho }_{+-,-+}^{g}\}^{\ast }=%
\widetilde{c}(s,\Theta _{t}) \,\, \frac{4p^{2}}{s} \,\, e^{i2\Phi
_{t}} \,\, \sin ^{4}\Theta _{t} \hskip 28mm {(C.2)} \nonumber
\end{eqnarray}
\begin{eqnarray}
\widetilde{\rho }_{-+,++}^{g} &=&\widetilde{\rho }_{-+,--}^{g}=-\widetilde{%
\rho }_{--,+-}^{g}=-\widetilde{\rho }_{++,+-}^{g}=  \hskip 27mm {(C.3)} \nonumber \\
\{\widetilde{\rho }_{++,-+}^{g}\}^{\ast } &=&\{\widetilde{\rho }%
_{--,-+}^{g}\}^{\ast }=-\{\widetilde{\rho }_{+-,--}^{g}\}^{\ast }=-\{%
\widetilde{\rho }_{+-,++}^{g}\}^{\ast }= \hskip 15mm {(C.4)} \nonumber \\
&&\widetilde{c}(s,\Theta _{t})\frac{8p^{2}m}{s^{3/2}} \,\,
e^{i\Phi _{t}} \,\, \sin ^{3}\Theta _{t}\cos \Theta _{t} \hskip
30mm {(C.5)} \nonumber
\end{eqnarray}
\begin{eqnarray}
\widetilde{\rho }_{++,++}^{g}=\widetilde{\rho }_{--,--}^{g}=\,\, \widetilde{c}%
(s,\Theta _{t}) \,\, \frac{4m^{2}}{s} \,\,
(1+\frac{4p^{2}}{s}[1+\sin ^{4}\Theta _{t}]) \hskip 25mm {(C.6)}
\nonumber
\end{eqnarray}
\begin{eqnarray}
\widetilde{\rho }_{++,--}^{g}=\widetilde{\rho }_{--,++}^{g}=\widetilde{c}%
(s,\Theta _{t}) \,\, \frac{4m^{2}}{s} \,\,
(1-\frac{4p^{2}}{s}[1+\sin ^{4}\Theta _{t}]) \hskip 25mm {(C.7)}
\nonumber
\end{eqnarray}
where
\begin{eqnarray}
\widetilde{c}(s,\Theta _{t})=\frac{  s^{2} g^{4} }{%
96(m^{2}-t)^{2}(m^{2}-u)^{2}} \,\, [7+\frac{36p^{2}}{s}\cos
^{2}\Theta _{t}] \hskip 28mm {(C.8)} \nonumber
\end{eqnarray}
Note that there is not an overall minus sign in (C.5) versus (32 )
in the text.  These ${ \widetilde{\rho}_{\lambda _{1}\lambda
_{2};\lambda _{1}^{^{\prime }}\lambda _{2}^{^{\prime }}}}$ were
derived using the $u(p,\pm )|_{0}, \ldots $  spinor outer-products
of Appendix A, see remarks after (A.2).  These gluon-production
density-matrix-elements (C.1-C.7) agree in magnitude with
\begin{eqnarray}
\frac{1}{4} \sum_{h_{g1},h_{g2}}(h_{g1},h_{g2},\lambda
_{1},\lambda _{2})(h_{g1},h_{g2},\lambda _{1}^{^{\prime }},\lambda
_{2}^{^{\prime }})^{\ast } \hskip 32mm {(C.9)} \nonumber
\end{eqnarray}
constructed from eq(A2) of reference [15].  The $e^{\pm i2 \Phi
_{t}}$, $e^{ \pm i\Phi _{t}}$ factors are missing in (C.9) and
some overall signs differ between (C.1-C.7) and (C.9).

From the derivation of $ { \widetilde{\rho}_{\lambda _{1}\lambda
_{2};\lambda _{1}^{^{\prime }}\lambda _{2}^{^{\prime }}}^{g}} $,
one obtains a simple ``Substitution Rule" for obtaining these
$\Theta _{t}$, $\Phi _{t}$ density-matrix-elements from the
$\Theta _{B}$, $\Phi _{B}$ ones in the text:
\newline $
\{$ $ \cos \Theta _{B} \rightarrow \cos \Theta _{t} $; $e^{i\Phi
_{B}}\rightarrow 1$; $\sin \Theta _{B}\rightarrow -e^{i\Phi
_{t}}\sin \Theta _{t}$ in ``${\lambda_1}$,${\lambda_2} = {-,+ }$"
and ``${{\lambda_1}^{'}}$, ${{\lambda_2}^{'}}$ $ = {+,-}$", with
complex-conjugation of this rule for ``$-$" $\leftrightarrow$
``$+$" cases $ \}$
\newline
This latter step in the substitution rule, which involves a minus
sign, occurs due to the sine functions which arise from $q\cdot
C,q\cdot B,\ldots $ factors in the spinor outer-products. This
latter step needs to be performed twice in some of the
helicity-conserving density-matrix-elements, and once in the
mixed-helicity density-matrix-elements.

The quark-production density-matrix-elements are
\begin{eqnarray}
\widetilde{\rho }_{+-,+-}^{q}=\widetilde{\rho }_{-+,-+}^{q}=\frac{g^{4}}{9}%
\,\, (1+\cos ^{2}\Theta _{t}) \hskip 25mm {(C.10)} \nonumber
\end{eqnarray}
\begin{eqnarray}
\widetilde{\rho }_{-+,+-}^{q}=\{\widetilde{\rho }_{+-,-+}^{q}\}^{\ast }=%
\frac{g^{4}}{9} \,\, e^{i2\Phi _{t}} \,\, \sin ^{2}\Theta _{t}
\hskip 42mm {(C.11)} \nonumber
\end{eqnarray}
\begin{eqnarray}
\widetilde{\rho }_{-+,++}^{q} &=&\widetilde{\rho }_{-+,--}^{q}=-\widetilde{%
\rho }_{--,+-}^{q}=-\widetilde{\rho }_{++,+-}^{q}=
\nonumber \\
\{\widetilde{\rho }_{++,-+}^{q}\}^{\ast } &=&\{\widetilde{\rho }%
_{--,-+}^{q}\}^{\ast }=-\{\widetilde{\rho }_{+-,--}^{q}\}^{\ast }=-\{%
\widetilde{\rho }_{+-,++}^{q}\}^{\ast }=
\nonumber \\
&&\frac{2mg^{4}}{9s^{1/2}} \,\, e^{i\Phi _{t}} \,\, \sin \Theta
_{t}\cos \Theta _{t} \hskip 37mm {(C.12)} \nonumber
\end{eqnarray}
\begin{eqnarray}
\widetilde{\rho }_{++,++}^{q}=\widetilde{\rho }_{--,--}^{q}=\widetilde{\rho }%
_{++,--}^{q}=\widetilde{\rho }_{--,++}^{q}=\frac{4m^{2}g^{4}}{9 s}
\,\, \sin ^{2}\Theta _{t} \hskip 25mm {(C.13)} \nonumber
\end{eqnarray}
There is not an overall minus sign in (C.12) versus (97) in the
text.  The above substitution rule also yields these $ {
\widetilde{\rho}_{\lambda _{1}\lambda _{2};\lambda _{1}^{^{\prime
}}\lambda _{2}^{^{\prime }}}^{q}} $ density-matrix-elements from
the  $ { {\rho}_{\lambda _{1}\lambda _{2};\lambda _{1}^{^{\prime
}}\lambda _{2}^{^{\prime }}}^{q}} $  ones in the text.
\end{appendix}

\newpage

\begin{center}
{\bf Figure Captions}
\end{center}

FIG. 1: The derivation of the general ``beam referenced
stage-two-spin-correlation" function begins in the ``home" or
starting coordinate system $ ( {x_h}, {y_h}, {z_h} ) $ in the $ (t
\bar{t} )_{c.m.}$ frame.  The top-quark $t_1$ is moving in the
positive $z_h$ direction, and $\theta_1, \phi_1$ specify the
${W_1}^+$ momentum direction. The $g_1$ gluon-momentum or $q_1$
quark-momentum ``beam" direction is specified by the spherical
angles $\Theta_B, \Phi_B$. Note that $\Phi_R = \Phi_B - \phi_1$.

FIG. 2:  Supplement to previous figure to show $\theta_2, \phi_2$
which specify the ${W_2}^-$ momentum direction.  The
azimuthal-angles $\phi_1$ and $\phi_2$ are Lorentz invariant under
boosts along the ${z_h}$ axis.  Note that the sum $\phi = \phi_1 +
\phi_2 $ is the angle between the $t_1$ and $\bar{t_2}$ decay
planes, and that the $\bar{t_2}$ momentum is in the negative $z_h$
direction. The three angles $\theta _1^t$, $\theta _2^t$ and $\phi
$ describe the first stage in the sequential-decays of the
$t\bar{t}$ system in which $ t_1 \rightarrow {W_1}^{+}b$ and
$\bar{t_2} \rightarrow {W_2}^{-}\bar b$.  The angles $\theta
_1^t$, $\theta _2^t$ are defined respectively in the $t_1$,
$\bar{t}_2$ rest frames, see [1].

FIG. 3: For derivation of the alternative $\Theta_t,\Phi_t$
beam-referenced production density-matrix-elements, a supplement
to Fig. 1 to show the $\Theta_t$ and $\Phi_t$ angles in the beam
coordinate system $ ( {x_b}, {y_b}, {z_b} ) $ in the $ (t \bar{t}
)_{c.m.}$ frame.  The $g_1$ gluon-momentum or $q_1$ quark-momentum
``beam" direction is in the positive $z_b$ direction. The final
$t_1$ momentum is specified by the spherical angles $\Theta_t,
\Phi_t$, with the $\bar{t_2}$ momentum, not shown, back to back
with the $t_1$.

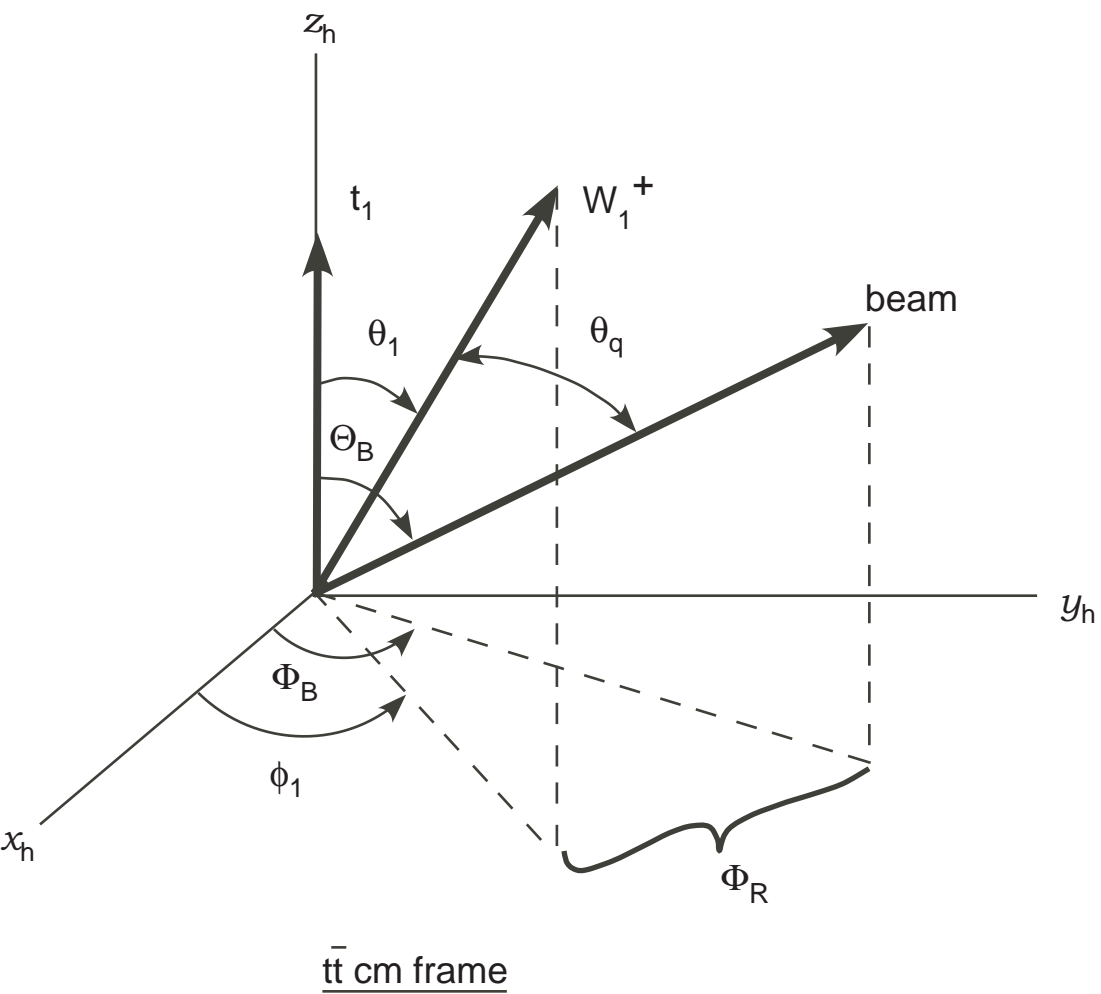
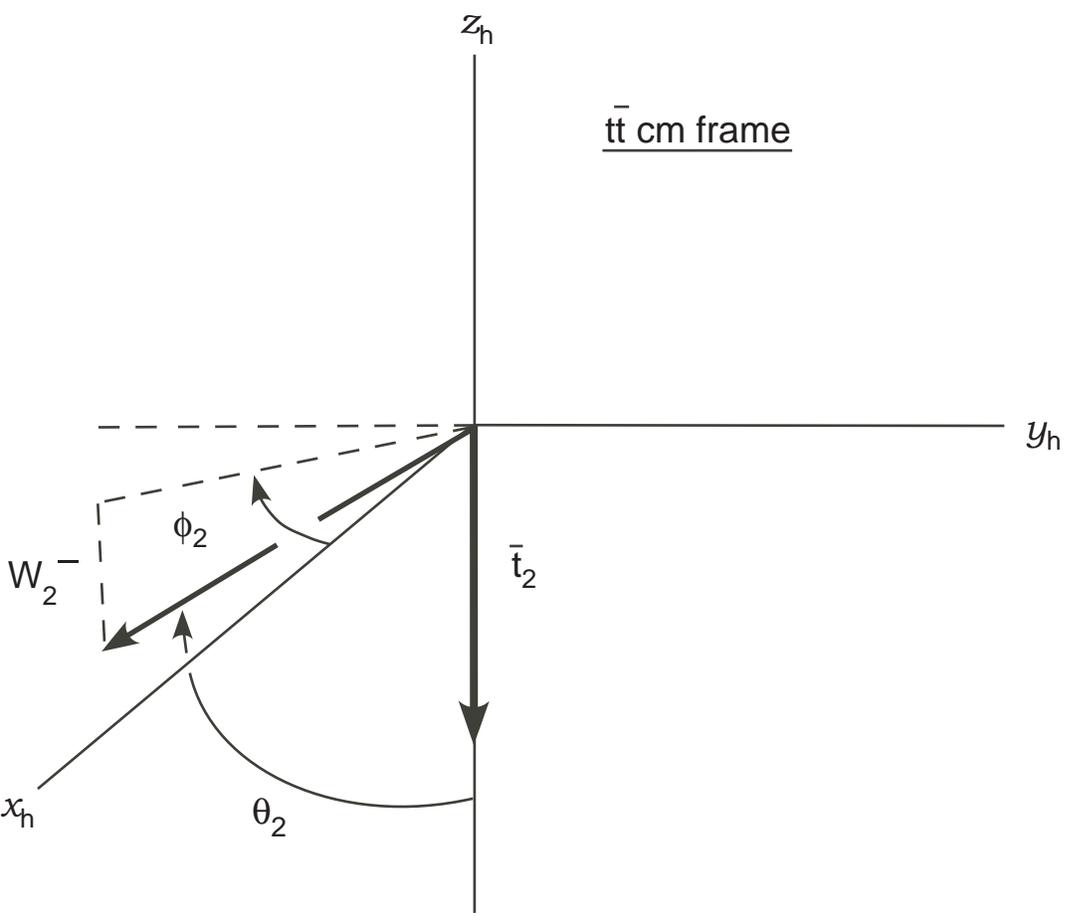
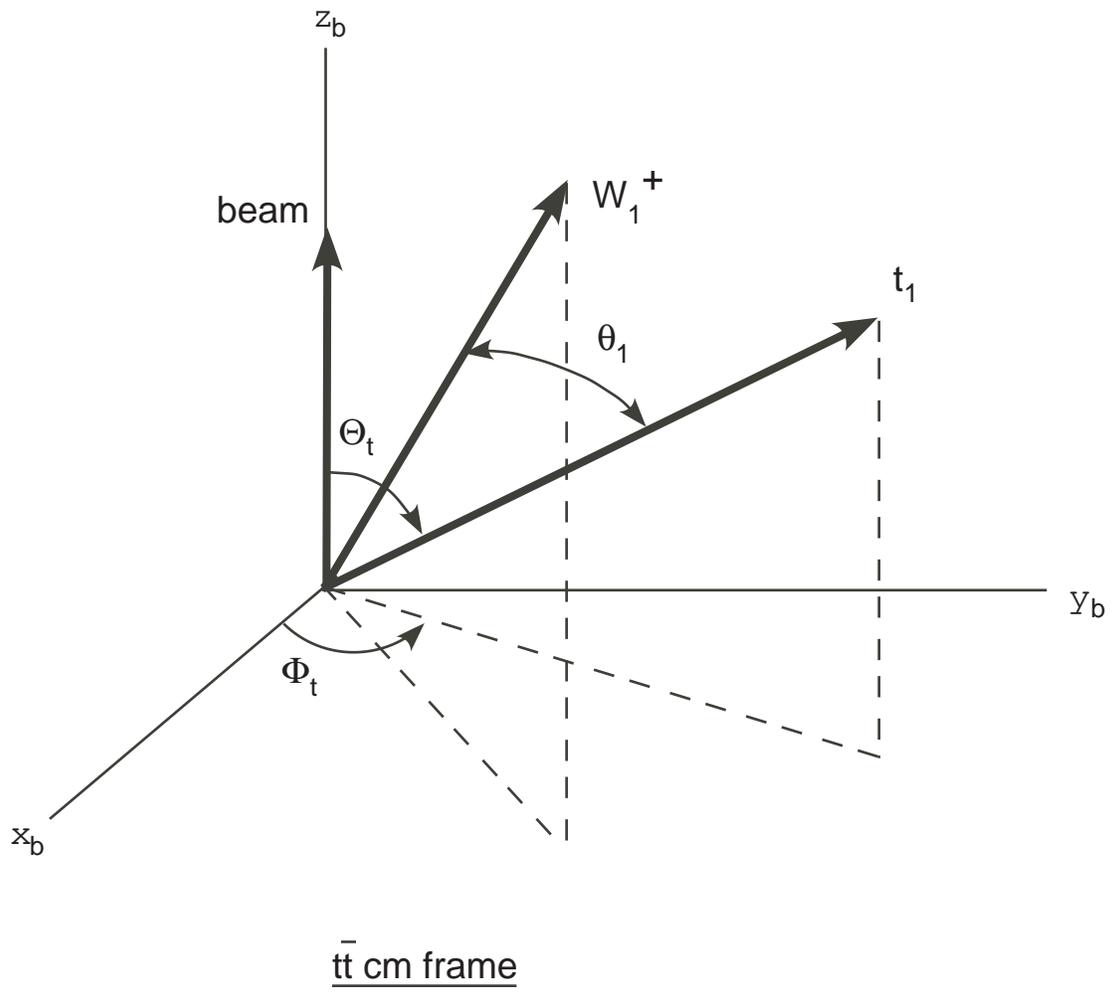
\end{document}